\documentclass[aps,prl,twocolumn,amsmath,superscriptaddress]{revtex4}  % for review and submission
\usepackage{bbm}
\usepackage{mathrsfs}
\usepackage{amsmath}
\usepackage{amsfonts}
\usepackage[colorlinks=true,citecolor=blue,anchorcolor=blue]{hyperref}
\usepackage{graphicx,epstopdf}
\usepackage{subfigure}
\usepackage{epsfig}
\usepackage{dcolumn}
\usepackage{bm}
\usepackage{color}
\usepackage{natbib}
\usepackage{amssymb}
\usepackage{xcolor}
\usepackage{braket}

\begin{document}

\title{Multiple topological nodal structure in LaSb$_2$ with large linear magnetoresistance}

\author{Y. X. Qiao}\altaffiliation{These authors contributed equally to this work}
\affiliation{State Key Laboratory of Functional Materials for Informatics, Shanghai Institute of Microsystem and Information Technology, Chinese Academy of Sciences, Shanghai 200050, China}
\affiliation{Center of Materials Science and Optoelectronics Engineering, University of Chinese Academy of Sciences, Beijing 100049, China}

\author{Z. C. Tao}\altaffiliation{These authors contributed equally to this work}
\affiliation{School of Physical Science and Technology, ShanghaiTech University, Shanghai 201210, China}

\author{F. Y. Wang}\altaffiliation{These authors contributed equally to this work}
\affiliation{National Laboratory of Solid State Microstructures, School of Physics, Nanjing University, Nanjing 210093, China}
\affiliation{Collaborative Innovation Center of Advanced Microstructures, Nanjing University, Nanjing 210093, China}

\author{Huaiqiang Wang}
\email{hqwang@nju.edu.cn}
\affiliation{National Laboratory of Solid State Microstructures, School of Physics, Nanjing University, Nanjing 210093, China}
\affiliation{Collaborative Innovation Center of Advanced Microstructures, Nanjing University, Nanjing 210093, China}

\author{Z. C. Jiang}
\affiliation{State Key Laboratory of Functional Materials for Informatics, Shanghai Institute of Microsystem and Information Technology, Chinese Academy of Sciences, Shanghai 200050, China}

\author{Z. T. Liu}
\affiliation{State Key Laboratory of Functional Materials for Informatics, Shanghai Institute of Microsystem and Information Technology, Chinese Academy of Sciences, Shanghai 200050, China}
\affiliation{Center of Materials Science and Optoelectronics Engineering, University of Chinese Academy of Sciences, Beijing 100049, China}

\author{Soohyun Cho}
\affiliation{State Key Laboratory of Functional Materials for Informatics, Shanghai Institute of Microsystem and Information Technology, Chinese Academy of Sciences, Shanghai 200050, China}

\author{F. Y. Zhang}
\affiliation{State Key Laboratory of Functional Materials for Informatics, Shanghai Institute of Microsystem and Information Technology, Chinese Academy of Sciences, Shanghai 200050, China}
\affiliation{Shenzhen Institute for Quantum Science and Technology and Department of Physics, Southern University of Science and Technology, Shenzhen 518055, China}

\author{Q. K. Meng}
\affiliation{Wuhan National High Magnetic Field Center and School of Physics, Huazhong University of Science and Technology, Wuhan 430074, China}

\author{W. Xia}
\affiliation{School of Physical Science and Technology, ShanghaiTech University, Shanghai 201210, China}
\affiliation{ShanghaiTech Laboratory for Topological Physics, ShanghaiTech University, Shanghai 201210, China}

\author{Y. C. Yang}
\affiliation{State Key Laboratory of Functional Materials for Informatics, Shanghai Institute of Microsystem and Information Technology, Chinese Academy of Sciences, Shanghai 200050, China}

\author{Z. Huang}
\affiliation{State Key Laboratory of Functional Materials for Informatics, Shanghai Institute of Microsystem and Information Technology, Chinese Academy of Sciences, Shanghai 200050, China}
\affiliation{School of Physical Science and Technology, ShanghaiTech University, Shanghai 201210, China}

\author{J. S. Liu}
\affiliation{State Key Laboratory of Functional Materials for Informatics, Shanghai Institute of Microsystem and Information Technology, Chinese Academy of Sciences, Shanghai 200050, China}
\affiliation{Center of Materials Science and Optoelectronics Engineering, University of Chinese Academy of Sciences, Beijing 100049, China}

\author{Z. H. Liu}
\affiliation{State Key Laboratory of Functional Materials for Informatics, Shanghai Institute of Microsystem and Information Technology, Chinese Academy of Sciences, Shanghai 200050, China}
\affiliation{Center of Materials Science and Optoelectronics Engineering, University of Chinese Academy of Sciences, Beijing 100049, China}

\author{Z. W. Zhu}
\affiliation{Wuhan National High Magnetic Field Center and School of Physics, Huazhong University of Science and Technology, Wuhan 430074, China}

\author{S. Qiao}
\affiliation{State Key Laboratory of Functional Materials for Informatics, Shanghai Institute of Microsystem and Information Technology, Chinese Academy of Sciences, Shanghai 200050, China}
\affiliation{Center of Materials Science and Optoelectronics Engineering, University of Chinese Academy of Sciences, Beijing 100049, China}
\affiliation{School of Physical Science and Technology, ShanghaiTech University, Shanghai 201210, China}

\author{Y. F. Guo}
\email{guoyf@shanghaitech.edu.cn}
\affiliation{School of Physical Science and Technology, ShanghaiTech University, Shanghai 201210, China}
\affiliation{ShanghaiTech Laboratory for Topological Physics, ShanghaiTech University, Shanghai 201210, China}

\author{Haijun Zhang}
\email{zhanghj@nju.edu.cn}
\affiliation{National Laboratory of Solid State Microstructures, School of Physics, Nanjing University, Nanjing 210093, China}
\affiliation{Collaborative Innovation Center of Advanced Microstructures, Nanjing University, Nanjing 210093, China}

\author{Dawei Shen}
\email{dwshen@mail.sim.ac.cn}
\affiliation{State Key Laboratory of Functional Materials for Informatics, Shanghai Institute of Microsystem and Information Technology, Chinese Academy of Sciences, Shanghai 200050, China}
\affiliation{Center of Materials Science and Optoelectronics Engineering, University of Chinese Academy of Sciences, Beijing 100049, China}

\date{\today}% It is always \today, today,

\begin{abstract}

Unconventional fermions in the immensely studied topological semimetals are the source for rich exotic topological properties. Here, using symmetry analysis and first-principles calculations, we propose the coexistence of multiple topological nodal structure in LaSb$_2$, including topological nodal surfaces, nodal lines and in particular eightfold degenerate nodal points, which have been scarcely observed in a single material. Further, utilizing high resolution angle-resolved photoemission spectroscopy in combination with Shubnikov-de Haas quantum oscillations measurements, we confirm the existence of nodal surfaces and eightfold degenerate nodal points in LaSb$_2$, and extract the $\pi$ Berry phase proving the non-trivial electronic band structure topology therein. The intriguing multiple topological nodal structure might play a crucial role in giving rise to the large linear magnetoresistance. Our work renews the insights into the exotic topological phenomena in LaSb$_2$ and its analogous. 

\end{abstract}

\maketitle

%\tableofcontents

Since the proposal of Weyl semimetal in 2011~\cite{Wan2011}, various kinds of topological semimetals (TSMs) characterized by gapless band touching nodes have been discovered and continuously attracted extensive research interest~\cite{Weng2016Topological, Liu2014Na3Bi, Lv2015Experimental, Soluyanov2015TypeII, Xu2015Discovery}. Generally speaking, three types of nodes can occur in TSMs, namely, nodal points~\cite{Armitage2018}, nodal lines~\cite{Burkov2011Topological, Fang2016Topological, Yu2017Topological, Yang2018Symmetry, Song2020Photoemission, Liu2018Experimental, Bzdusek2016Nodal, Yan2018Experimental}, and nodal surfaces~\cite{Wu2018Nodal, Fu2019Dirac, Song2022Spectroscopic}. These nodal states in TSMs manifest themselves not only in exotic surface states, such as Fermi-arc~\cite{Xu2015Observation, Rao2019Observation} and drumhead surface states~\cite{Bian2016Topological, Schoop2016Dirac, Fu2019Dirac}, but also in unique transport behaviors, e.g., the chiral negative magnetoresistance (MR)~\cite{Ali2014Large,Huang2015Observation,Li2016Negative}, and the chiral magnetic effect~\cite{Fukushima2008Chiral, Schroter2019Chiral}. Intriguingly, it has been shown that nonsymmorphic symmetries can give rise to many fancy degeneracies in band structures, say, unconventional quasiparticles with higher-order degeneracies beyond Dirac and Weyl fermions~\cite{Bradlyn2016Beyond, Guo2021Eightfold, Sun2017Coexistence, Tang2017Multiple, Wu2021Symmetry, Xia2017Triply, Zhang2017Coexistence}. To date, both three- and six-fold degenerate points (DPs) have been confirmed experimentally~\cite{Lv2017Observation,Ma2018Threecomponent,Rao2019Observation,Sun2020Direct}. Whereas, unambiguous experimental observations of eightfold DPs is still rare~\cite{Schoop2018Tunable, Berry2020Laser}, partly because of the lack of realistic materials hosting eightfold DPs near the Fermi level.

Recently, rare earth diantimonide LaSb$_2$ has aroused great interest, since it exhibits charge density wave\cite{2022arXiv220203161P, 2022arXiv220804997B}, superconductivity~\cite{Guo2011Dimensional, Galvis2013Scanning} and anomalous transport properties~\cite{Bud1998Anisotropic, Goodrich2004deHaasvan, Young2003High}. More importantly, owning to its large linear magnetoresistance (LMR) ($\sim$10000\%), it has been viewed as a compelling candidate for high-field sensors~\cite{Goodrich2004deHaasvan}. However, so far the origin of the LMR in LaSb$_2$ still remains unknown, even though there have been some explanations based on classical~\cite{Parish2003Nonsaturating,Wang2012Room,Dasoundhi2021Extremely,Qu2010Quantum} and quantum effects~\cite{Rusza2020DiracLike}, respectively. As a nonsymmorphic crystal, LaSb$_2$ holds promises for harboring topological nodal states~\cite{Leonhardt2021}, which has been previously largely overlooked. Further, considering the intimate relationship between gapless Dirac-like nodal states and 
LMR~\cite{Abrikosov1998Quantum}, more in-depth theoretical and experimental studies on LaSb$_2$ are thus highly desired. 

\begin{figure*}
\includegraphics {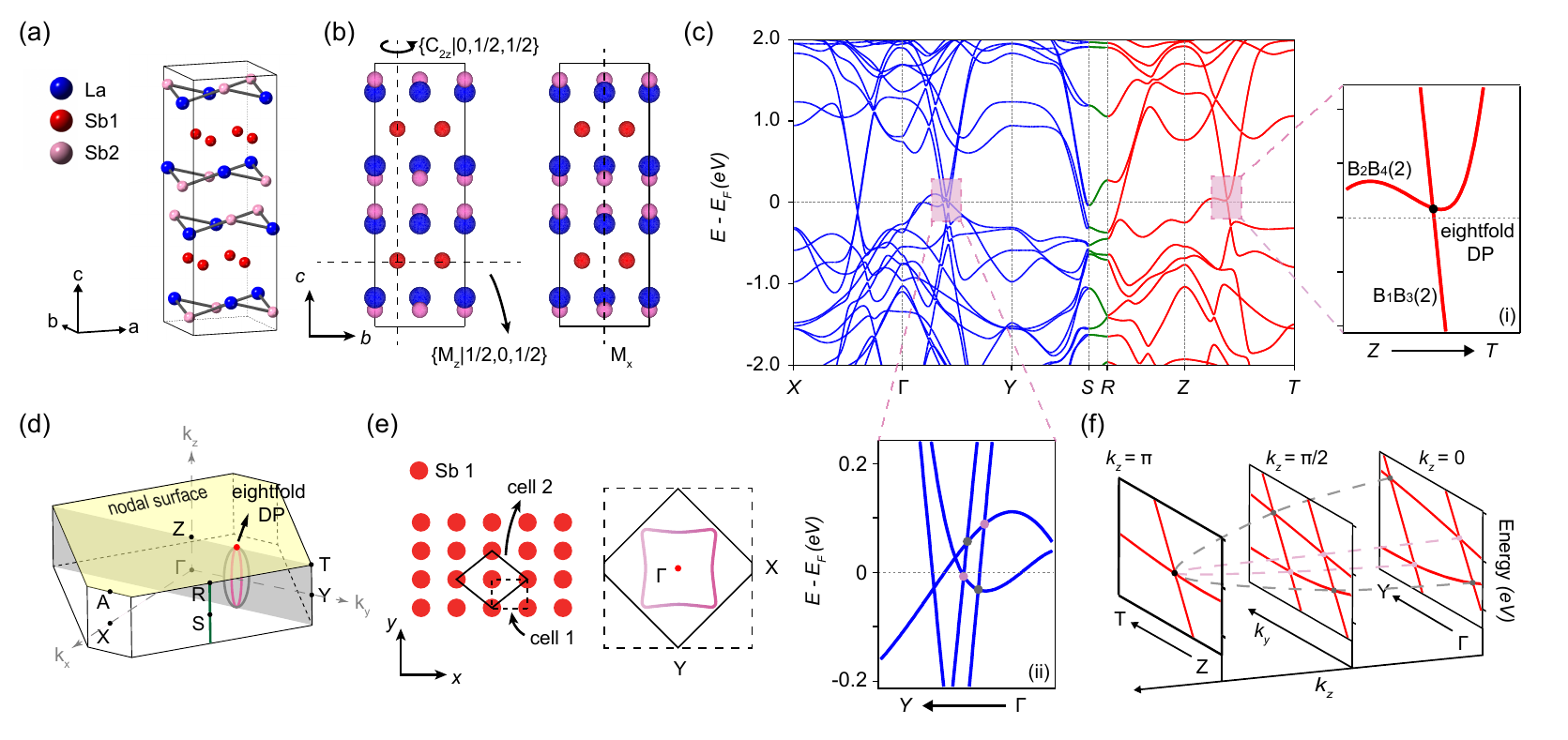}
\caption{\label{calculation} Electronic structures of LaSb$_2$. (a) 3D crystal structure of LaSb$_2$. (b)  Some schematic important crystal symmetries of LaSb$_2$. (c) Calculated electronic band structures without SOC. Blue (green) and red lines represent double and fourfold degeneracies, respectively, where spin degeneracy is included. (i) and (ii) Enlarged views of the band structure without SOC along high-symmetry lines in the $k_x$=0 plane, corresponding to pink boxes in (c). The eightfold DP (counting spin degeneracy) exists along the $Z$-$T$ line. For each spin, it results from the crossing of two sets of doubly degenerate bands with different two-dimensional irreducible representations $B_1 B_3$ and $B_2 B_4$, respectively. (d) Schematic of various nodal characters of LaSb$_2$ in the bulk BZ. Yellow and grey planes represent the $k_z=\pi$ plane (nodal surface), and $k_x=0$ plane. (e) Schematic of diamond-like nodal loop (pink line) in the $k_z=0$ plane from BZ-folding process for LaSb$_2$ with slightly distorted square nets of Sb1 atoms. The dashed (solid) lines denote the BZ before (after) folding procedure. (f) Schematic of the splitting of the eightfold DP from the $k_z=\pi$ line to general $k_z$ values in the $k_x=0$ plane.}
\end{figure*}

In this Letter, with a combination of symmetry analysis and first-principles calculation, we propose a plethora of topological nodal structures in LaSb$_2$, covering all three types of nodes. Notably, we uncover symmetry-protected eightfold DPs on the nodal surface near the Fermi level. Using high-resolution angle-resolved photoemission spectroscopy (ARPES), we unambiguously confirm the predicted nodal surface and most importantly, the eightfold DPs. Our further transport measurements reveal the imperfect carriers compensation in LaSb$_2$~\cite{Young2003High,Rusza2020DiracLike}, and a nontrivial $\pi$ Berry phase from the quantum oscillation analysis. Our findings suggest that the LMR observed in LaSb$_2$ should be attributed to Dirac nodal states rather than the electron-hole compensation mechanism. In addition, given that LaSb$_2$ shows a unique superconducting transition ($T_c = 1.2$ K) among all rare earth diantimonides, it might provide an ideal platform to explore the interplay between symmetry, non-trivial band topology and superconductivity in systems with multiple topological nodal states.

\begin{figure*}
\includegraphics {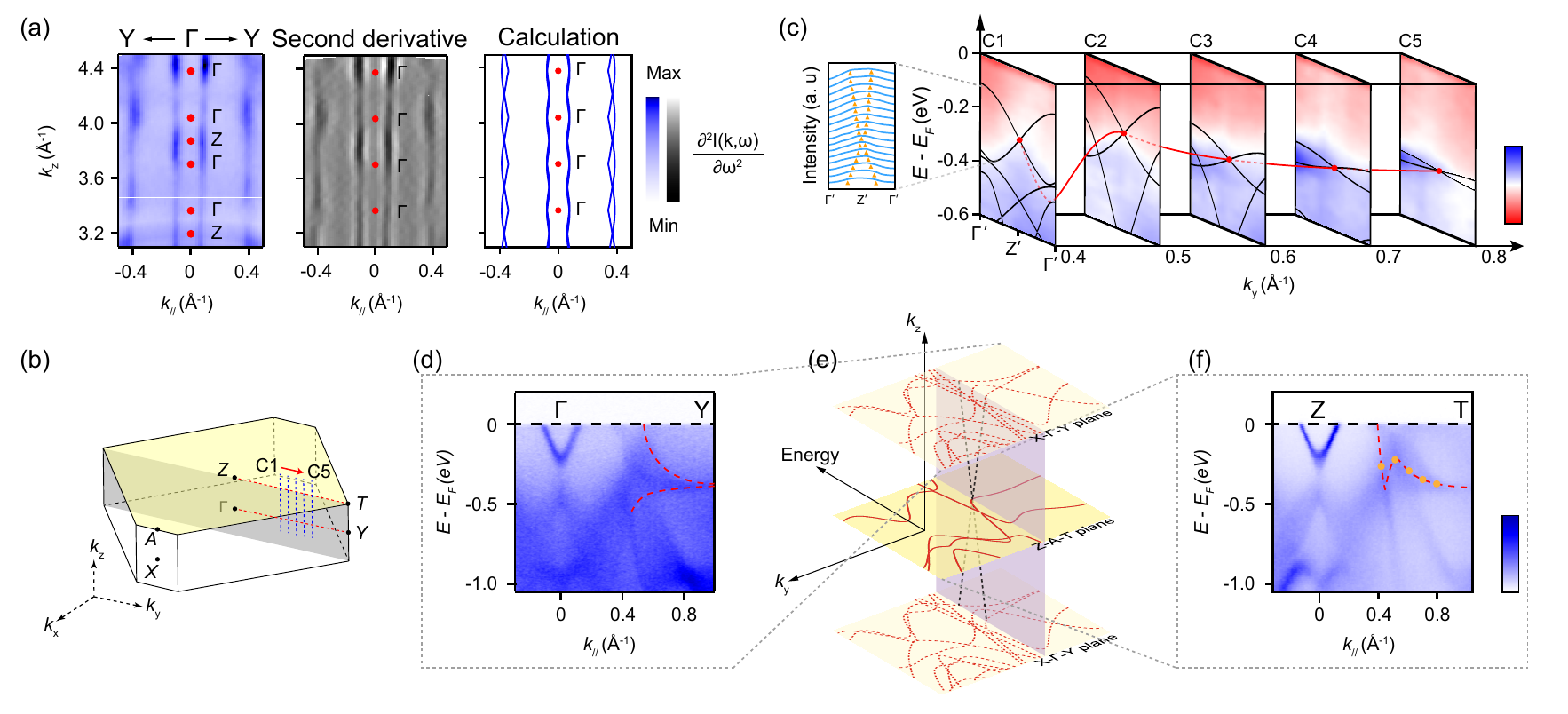}
\caption{\label{nodalsurface} Verification of nodal surface in LaSb$_2$. (a) Integrated photoemission intensity map in the $k_y-k_z$ plane taken at $E_F$, corresponding second-derivative plot and calculation result. (b) Schematic of momentum locations of cuts C1 to C5, $Z$-$T$ and $\Gamma$-$Y$ in the bulk BZ. (c) Schematic 3D plot of the electronic structure of C1 to C5. The band dispersion along $k_z$ direction probed by different photon energies and corresponding calculations plotted by the black curves. The red line illustrates the four-fold degenerate band shown in (f). Momentum distribution curves of C1 plotted at left bottom. (e) Schematic plot of the band structure in $k_z=0$ and nodal surface plane. Thin and dotted red curves represent the degenerate band on the nodal surface and the non-degenerate bands away from the nodal surfaces, respectively. (d) and (f) The band structure dispersion along $\Gamma-Y$ and $Z$-$T$ directions and the corresponding calculation results as shown in top and middle of (e), respectively. Typical non-degenerated and double degenerated band are highlighted by red dots line.} 
\end{figure*}

LaSb$_2$ belongs to the RSb$_2$ (R = La-Nd, and Sm) family which crystallize in the orthorhombic, highly layered SmSb$_2$-like structure. Here, alternating La/Sb layers and two-dimensional rectangular sheets of Sb atoms are stacked along the $c$ axis~\cite{Fischer2019Transport}, as displayed in Fig.~\ref{calculation}(a). Details on the methods of sample growth, characterization, ARPES and calculation are described in Note \uppercase\expandafter{\romannumeral1} of Supplementary Material (SM)~\cite{SM}. We note that our calculations mainly focus on the result without considering the spin-orbit coupling (SOC) because of the relatively weak SOC in LaSb$_2$. Unless otherwise specified, henceforth spin degeneracy has been assumed. The case with the SOC is discussed in Note \uppercase\expandafter{\romannumeral2} of SM~\cite{SM}. 

Based on the symmetry analysis, nodal surface states with fourfold degeneracy inevitably emerge in the band structure in the $k_z=\pi$ plane [shaded in yellow in Fig.~\ref{calculation}(d)], as validated by the red-colored bands along the $R$-$Z$-$T$ path in Fig.~\ref{calculation}(c). Such nodal surfaces stem from the Kramer's degeneracy imposed by the combined antiunitary symmetry operator $S_{2z}\Theta$ in the $k_z=\pi$ plane with $(S_{2z}\Theta)^{2}=-1$ [see more details in Note \uppercase\expandafter{\romannumeral1} of SM~\cite{SM}], where $S_{2z}=\{C_{2z}\ |\ 0,1/2,1/2\}$ [Fig.~\ref{calculation}(b)] and $\Theta$ denote the screw rotation and time-reversal operators, respectively. It is worth mentioning that the similar Kramer's degeneracy gives rise to fourfold degenerate straight nodal lines along $R$-$S$ direction~\cite{SM}, where $(G_z\Theta)^2=-1$ with $G_z=\{m_z\ |\ 0, 1/2,1/2\}$ representing the glide symmetry operation, as shown in Fig.\ref{calculation}(b).
 
%See Supplemental Material (SM)~\cite{SM} and references therein~\cite{SM} for material, characterization and calculation details.

Intriguingly, two eightfold DPs related to each other by the time-reversal symmetry, from the crossing of two set of nodal surface bands, appear in the high symmetry $Z$-$T$ line, as can be seen in Fig.~\ref{calculation}(c) and its zoomed-in inset (i). Each eightfold DP consists of two fourfold DPs from the two spin species. For each spin, the fourfold DP results from the crossing between two set of doubly degenerate bands with distinct two-dimensional irreducible representations along the $Z$-$T$ direction, which are labelled by $B_1 B_3$ and $B_2 B_4$, respectively. One notable feature of the eightfold DPs in LaSb$_2$ is that they reside rather close to the Fermi energy ($E_F$), thus enabling their experimental observation and further potential application. Moreover, away from the $Z$-$T$ line in the $k_x=0$ plane, due to the reduced symmetry, each eightfold DP would split into four nodal points, as shown in the lower zoomed-in inset (ii) of Fig.~\ref{calculation}(c). Intriguingly, these nodal points form two nodal lines in the $k_x=0$ plane [shaded in grey in Fig.~\ref{calculation}(d)], and are protected by the $M_x$ mirror symmetry schematically shown in Fig. ~\ref{calculation}(b). For a better clarification, in Fig. ~\ref{calculation}(f), by choosing three representative cuts along the $k_y$ direction with different $k_z$ values, we demonstrate how the two nodal lines merge into the nodal point.

In addition, there are Dirac-type linear bands within a large energy range (~2 eV) along $\Gamma$-$X$, which are dominated by the $p_{x/y}$ orbitals of Sb1 atoms [see Fig. S2(b) in SM~\cite{SM}]. Similar to the case of square-net materials~\cite{Klemenz2019}, such as ZrSiS~\cite{Schoop2016Dirac, Fu2019Dirac}, the Dirac-like bands result from the BZ folding of the slightly distorted square-net Sb1 atoms in the $\sqrt{2}\times\sqrt{2}$ supercell structure [see the left panel of Fig.~\ref{calculation}(e)]. Such Dirac-type crossings form a diamond-like nodal loop in the $k_z=0$ plane, as shown by pink lines in the right panel of Fig.~\ref{calculation}(e). 

\begin{figure*}
\includegraphics {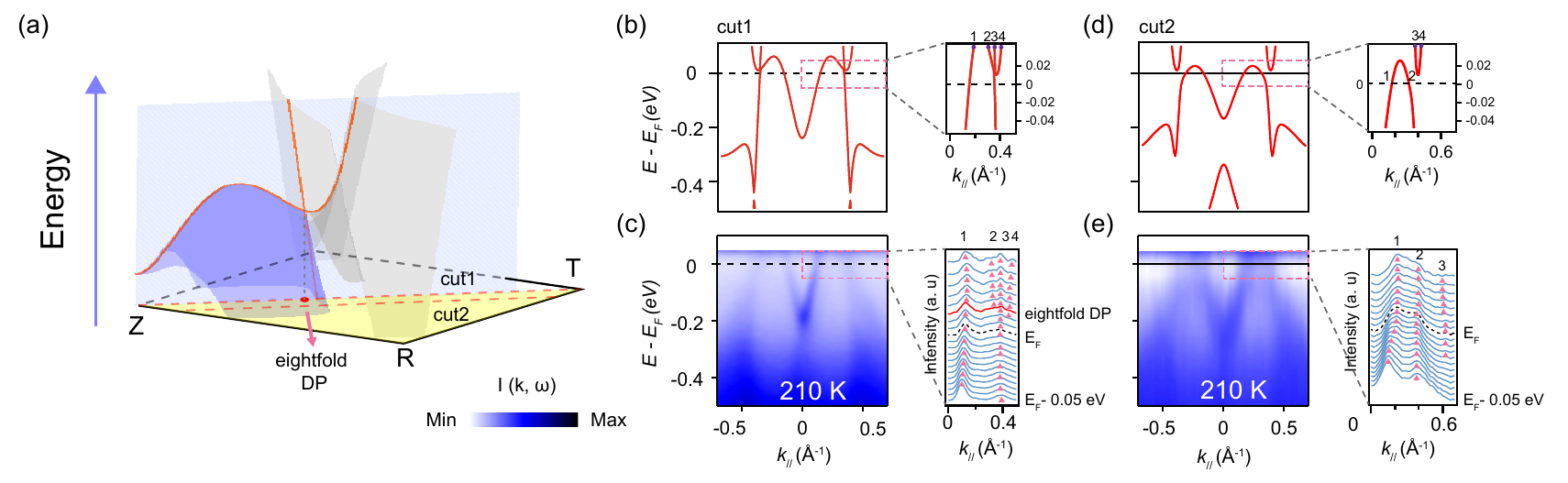}
\caption{\label{eightfold} Electronic structures of eightfold degenerate point of LaSb$_2$. (a) Schematic 3D plot of the calculated electronic structure near eightfold degenerate point in the $k_z=\pi$ plane. The red dots lines represent the location of cut 1 and cut 2, respectively. The yellow plane represent the nodal surface. (c) and (e) Photoemission intensity plot and corresponding momentum distribution curves (MDCs) of the region in the pink box in band structure at high temperature along cut1 and cut2, respectively. Cuts 1 and 2 are illustrated. The pink triangles in MDCs indicate the peak positions, which track the dispersions of the bands through the Fermi level. (b) and (d) The corresponding calculations of (c) and (d).}
\end{figure*}

To verify these predicted nodal surface states in LaSb$_2$, we firstly carried out photon-energy-dependent measurements to identify the $k_z$-dependent electronic structures. The left and middle panels of Fig.~\ref{nodalsurface}(a) show the photoemission intensity map taken at $E_F$ ($h\nu =28\sim70$~eV) and the corresponding second-derivative plot, respectively. One can distinguish the warped chainlike band feature with a pronounced periodic modulation along the $k_z$ direction, indicating that our measurements can probe the bulk states of LaSb$_2$. With an inner potential of 14 eV and $c = 18.671$ $\AA$, we fitted the periodic variation according to the free electron final-state mode~\cite{Liebowitz1978Free}, and we thus determined $h\nu = 75, 69 $~eV are close to $\Gamma$ and $Z$ point, respectively. The fitted result is in good agreement with the bulk band calculation [the right panel of Fig.~\ref{nodalsurface}(a)].

According to our calculations, nodal surfaces with fourfold degenerated band crossings are located in $k_z=\pi$ planes [Fig.~\ref{nodalsurface}(b)].
As the most pronounced fingerprint, any bands would show Dirac-like feature when crossing the nodal surface, and the degenerated band would split away from the nodal surface. Figure~\ref{nodalsurface}(c) presents photoemission intensity plots along cuts C1 to C5, which are all perpendicular to the nodal surface. For spectra along these cuts shown in Fig.~\ref{nodalsurface}(c), we can distinguish a series of Dirac-like band crossings through the corresponding momentum distribution curves (MDCs) (with the representative MDCs for C1 shown in the inset), which is as well roughly in line with the bulk calculation (black lines). We note that these crossing points (red dots) constitute the degenerate bands along $Z-T$ (the red dashed line) in Fig.~\ref{nodalsurface}(f). In sharp contrast, once a cut deviates from the $k_z=\pi$ plane, e.g., the cut along $\Gamma-Y$ in the $k_z=0$ plane, the originally degenerated band would split, as illustrated in Fig.~\ref{nodalsurface}(d), which is in good agreement with our calculation [Fig.~\ref{nodalsurface}(e)]. Thus, we can conclude that Dirac-like band crossings on the $k_z=0$ plane are protected by the Kramer's degeneracy, leading to the formation of the nodal surface.

According to our symmetry analysis and first-principles calculation, there should exist accidental crossings between two degenerate bands along the $T-Z-T$ high symmetry direction, which results in the formation of eightfold degenerate nodal point. However, this predicted degenerate point is suggested to be located slightly above $E_F$, beyond the probing scope of photoemission spectroscopy. To explore the unoccupied states, we performed relatively high temperature (210~K) ARPES measurements and the resulting data have been divided by the resolution convoluted Fermi-Dirac distribution function to highlight band dispersions above $E_F$. A schematic plot of the band structure near the eightfold degenerate point in the $Z-T$ line is displayed in Fig.~\ref{eightfold}(a). Along Cut~1, which would go through the eightfold degenerate nodal point, the calculated dispersion suggests the touching of the hole-(1,2 branches) and electron-like (3,4 branches) bands at the degenerate point, as shown in Fig.~\ref{eightfold}(b). Our photoemission intensity plot along Cut 1 shows general agreement with the calculation [Fig.~\ref{eightfold}(c)]. In particular, the corresponding MDCs for the zoomed-in region in the vicinity of $E_F$ could well resolve the touching point ($E_D$) of all the four band branches, strongly suggesting the existence of the nodal point. Given that all points on the $k_z$ plane should be fourfold degenerate, such a crossing point between two degenerate bands would be eightfold degenerate. Note that our data could not resolve the proposed gap owing to the SOC, which might be due to the relatively poor energy resolution of ARPES at such a high temperature. For comparison, we measured the band dispersion along Cut~2, which slightly deviates from Cut~1 but is parallel to that, as illustrated in Fig.~\ref{eightfold}(e). We found that there is no band crossing between band 2 and 3 within cut2 and its corresponding MDCs. Through careful comparison to the calculation [Fig.~\ref{eightfold}(d)], we could discover the splitting of nodal point between the hole- (1,2 branches) and electron-like (3,4 branches) bands. In this way, we experimentally verify the existence of eightfold degenerate nodal point on the nodal surface of LaSb$_2$.

Magneto-transport measurements were also performed on LaSb$_{2} $ to explore the topological nature of electronic bands which are located rather close to $E_F$. As shown in Note \uppercase\expandafter{\romannumeral3} of SM~\cite{SM}, the magneto-transport measurements unveil clear quantum oscillations at $B$ \textgreater 30 T below 10 K and exhibit striking Shubnikov–de Haas (SdH) oscillations after subtracting a sixth-order polynomial background, as show in SM~\cite{SM} Fig. S4. The Berry phase $ \varphi_{B} $ was examined by using the Landau level (LL) index fan, which gives the Berry phases $ \varphi_{B} $ of 0.17$ \pi $ and 1.01$ \pi $ for the electronic pocket $ \alpha $ and hole pocket $ \beta $, respectively. The Berry phase for the $ \beta $ frequency is close to $ \pi $, thus demonstrating the nontrivial topological state in LaSb$_2$.

As we mentioned above, LaSb$_2$ hosts unconventional LMR~\cite{Goodrich2004deHaasvan,Young2003High}, while its origin still remains elusive. %Because electron-hole (e-h) compensation often can give rise to unsaturated large magnetoresistance in majority of topological semimetals. 
To experimentally examine this, we conducted both magnetic field dependent Hall conductivity and ARPES measurements on LaSb$_2$ to estimate the carrier density of holes and electrons (see details in Notes \uppercase\expandafter{\romannumeral3} and \uppercase\expandafter{\romannumeral4} of SM~\cite{SM}). We found that there exists a huge difference in the hole and electron carrier density in LaSb$_2$, which unambiguously excludes the carriers compensation mechanism for the LMR therein. Our finding is as well consistent with the previous theoretical research~\cite{Rusza2020DiracLike}. Instead, we suggest that the quantum limit of Dirac-like nodal fermions with high Fermi velocities in high fields might play a more crucial role in resulting the unconventional LMR.

In summary, by combining first-principles calculations and ARPES measurements, as well as Shubnikov–de Haas oscillations, we unambiguously confirmed the existence of nodal surface and eightfold degenerate nodal points in the vicinity of $E_F$ in LaSb$_2$. The discovery of multiple nodal phenomena in this single material enables further identification of abundant topological properties in this old light rare-earth diantimonides system, making it a versatile platform for studying the interplay between topological characters and LMR.

%we have revealed the quantum effect is the main reason of large, linear MR. It is important for further theoretical developing and practical applications about magnetic devices. In addition, we have first identify that LaSb$_2$ is a Dirac nodal semimetal and it possesses abundant topological properties. 
\begin{acknowledgements}
We thank D. X. Shao and Prof. Gang Mu for the useful discussion. This work was supported by the the National Science Foundation of China (Grants No. U2032208, 12222413, 11874264, 12074181, 11834006 and 12104217) and the Natural Science Foundation of Shanghai (Grants No. 22ZR1473300 and 14ZR1447600). Y. F. Guo acknowledges the Shanghai Science and Technology Innovation Action Plan (Grant No. 21JC1402000) and the Open Project of Key Laboratory of Artificial Structures and Quantum Control (Ministry of Education), Shanghai Jiao Tong University (Grant No. 2020-04). JSL thanks the fund of Science and Technology on Surface Physics and Chemistry Laboratory (6142A02200102). Part of this research used Beamline 03U of the Shanghai Synchrotron Radiation Facility, which is supported by ME$^2$ project under contract No. 11227902 from National Natural Science Foundation of China. Haijun Zhang is supported by National Key Projects for Research and Development of China (Grant No.2021YFA1400400), the Fundamental Research Funds for the Central Universities (Grant No. 020414380185), Natural Science Foundation of Jiangsu Province (No. BK20200007) and the Fok Ying-Tong Education Foundation of China (Grant No. 161006).

\end{acknowledgements}
\bibliography{ref}

\begin{thebibliography}{61}
\expandafter\ifx\csname natexlab\endcsname\relax\def\natexlab#1{#1}\fi
\expandafter\ifx\csname bibnamefont\endcsname\relax
  \def\bibnamefont#1{#1}\fi
\expandafter\ifx\csname bibfnamefont\endcsname\relax
  \def\bibfnamefont#1{#1}\fi
\expandafter\ifx\csname citenamefont\endcsname\relax
  \def\citenamefont#1{#1}\fi
\expandafter\ifx\csname url\endcsname\relax
  \def\url#1{\texttt{#1}}\fi
\expandafter\ifx\csname urlprefix\endcsname\relax\def\urlprefix{URL }\fi
\providecommand{\bibinfo}[2]{#2}
\providecommand{\eprint}[2][]{\url{#2}}

\bibitem[{\citenamefont{Wan et~al.}(2011)\citenamefont{Wan, Turner, Vishwanath,
  and Savrasov}}]{Wan2011}
\bibinfo{author}{\bibfnamefont{X.}~\bibnamefont{Wan}},
  \bibinfo{author}{\bibfnamefont{A.~M.} \bibnamefont{Turner}},
  \bibinfo{author}{\bibfnamefont{A.}~\bibnamefont{Vishwanath}},
  \bibnamefont{and} \bibinfo{author}{\bibfnamefont{S.~Y.}
  \bibnamefont{Savrasov}}, \bibinfo{journal}{Phys. Rev. B}
  \textbf{\bibinfo{volume}{83}}, \bibinfo{pages}{205101}
  (\bibinfo{year}{2011}).

\bibitem[{\citenamefont{Weng et~al.}(2016)\citenamefont{Weng, Dai, and
  Fang}}]{Weng2016Topological}
\bibinfo{author}{\bibfnamefont{H.}~\bibnamefont{Weng}},
  \bibinfo{author}{\bibfnamefont{X.}~\bibnamefont{Dai}}, \bibnamefont{and}
  \bibinfo{author}{\bibfnamefont{Z.}~\bibnamefont{Fang}}, \bibinfo{journal}{J.
  Phys. Condens. Matter} \textbf{\bibinfo{volume}{28}}, \bibinfo{pages}{303001}
  (\bibinfo{year}{2016}).

\bibitem[{\citenamefont{Liu et~al.}(2014)\citenamefont{Liu, Zhou, Zhang, Wang,
  Weng, Prabhakaran, Mo, Shen, Fang, Dai et~al.}}]{Liu2014Na3Bi}
\bibinfo{author}{\bibfnamefont{Z.~K.} \bibnamefont{Liu}},
  \bibinfo{author}{\bibfnamefont{B.}~\bibnamefont{Zhou}},
  \bibinfo{author}{\bibfnamefont{Y.}~\bibnamefont{Zhang}},
  \bibinfo{author}{\bibfnamefont{Z.~J.} \bibnamefont{Wang}},
  \bibinfo{author}{\bibfnamefont{H.~M.} \bibnamefont{Weng}},
  \bibinfo{author}{\bibfnamefont{D.}~\bibnamefont{Prabhakaran}},
  \bibinfo{author}{\bibfnamefont{S.~K.} \bibnamefont{Mo}},
  \bibinfo{author}{\bibfnamefont{Z.~X.} \bibnamefont{Shen}},
  \bibinfo{author}{\bibfnamefont{Z.}~\bibnamefont{Fang}},
  \bibinfo{author}{\bibfnamefont{X.}~\bibnamefont{Dai}}, \bibnamefont{et~al.},
  \bibinfo{journal}{Science} \textbf{\bibinfo{volume}{343}},
  \bibinfo{pages}{864} (\bibinfo{year}{2014}).

\bibitem[{\citenamefont{Lv et~al.}(2015)\citenamefont{Lv, Weng, Fu, Wang, Miao,
  Ma, Richard, Huang, Zhao, Chen et~al.}}]{Lv2015Experimental}
\bibinfo{author}{\bibfnamefont{B.~Q.} \bibnamefont{Lv}},
  \bibinfo{author}{\bibfnamefont{H.~M.} \bibnamefont{Weng}},
  \bibinfo{author}{\bibfnamefont{B.~B.} \bibnamefont{Fu}},
  \bibinfo{author}{\bibfnamefont{X.~P.} \bibnamefont{Wang}},
  \bibinfo{author}{\bibfnamefont{H.}~\bibnamefont{Miao}},
  \bibinfo{author}{\bibfnamefont{J.}~\bibnamefont{Ma}},
  \bibinfo{author}{\bibfnamefont{P.}~\bibnamefont{Richard}},
  \bibinfo{author}{\bibfnamefont{X.~C.} \bibnamefont{Huang}},
  \bibinfo{author}{\bibfnamefont{L.~X.} \bibnamefont{Zhao}},
  \bibinfo{author}{\bibfnamefont{G.~F.} \bibnamefont{Chen}},
  \bibnamefont{et~al.}, \bibinfo{journal}{Phys. Rev. X}
  \textbf{\bibinfo{volume}{5}}, \bibinfo{pages}{031013} (\bibinfo{year}{2015}).

\bibitem[{\citenamefont{Soluyanov et~al.}(2015)\citenamefont{Soluyanov, Gresch,
  Wang, Wu, Troyer, Dai, and Bernevig}}]{Soluyanov2015TypeII}
\bibinfo{author}{\bibfnamefont{A.~A.} \bibnamefont{Soluyanov}},
  \bibinfo{author}{\bibfnamefont{D.}~\bibnamefont{Gresch}},
  \bibinfo{author}{\bibfnamefont{Z.}~\bibnamefont{Wang}},
  \bibinfo{author}{\bibfnamefont{Q.}~\bibnamefont{Wu}},
  \bibinfo{author}{\bibfnamefont{M.}~\bibnamefont{Troyer}},
  \bibinfo{author}{\bibfnamefont{X.}~\bibnamefont{Dai}}, \bibnamefont{and}
  \bibinfo{author}{\bibfnamefont{B.~A.} \bibnamefont{Bernevig}},
  \bibinfo{journal}{Nature} \textbf{\bibinfo{volume}{527}},
  \bibinfo{pages}{495} (\bibinfo{year}{2015}).

\bibitem[{\citenamefont{Xu et~al.}(2015{\natexlab{a}})\citenamefont{Xu,
  Belopolski, Alidoust, Neupane, Bian, Zhang, Sankar, Chang, Yuan, Lee
  et~al.}}]{Xu2015Discovery}
\bibinfo{author}{\bibfnamefont{S.~Y.} \bibnamefont{Xu}},
  \bibinfo{author}{\bibfnamefont{I.}~\bibnamefont{Belopolski}},
  \bibinfo{author}{\bibfnamefont{N.}~\bibnamefont{Alidoust}},
  \bibinfo{author}{\bibfnamefont{M.}~\bibnamefont{Neupane}},
  \bibinfo{author}{\bibfnamefont{G.}~\bibnamefont{Bian}},
  \bibinfo{author}{\bibfnamefont{C.}~\bibnamefont{Zhang}},
  \bibinfo{author}{\bibfnamefont{R.}~\bibnamefont{Sankar}},
  \bibinfo{author}{\bibfnamefont{G.}~\bibnamefont{Chang}},
  \bibinfo{author}{\bibfnamefont{Z.}~\bibnamefont{Yuan}},
  \bibinfo{author}{\bibfnamefont{C.~C.} \bibnamefont{Lee}},
  \bibnamefont{et~al.}, \bibinfo{journal}{Science}
  \textbf{\bibinfo{volume}{349}}, \bibinfo{pages}{613}
  (\bibinfo{year}{2015}{\natexlab{a}}).

\bibitem[{\citenamefont{Armitage et~al.}(2018)\citenamefont{Armitage, Mele, and
  Vishwanath}}]{Armitage2018}
\bibinfo{author}{\bibfnamefont{N.~P.} \bibnamefont{Armitage}},
  \bibinfo{author}{\bibfnamefont{E.~J.} \bibnamefont{Mele}}, \bibnamefont{and}
  \bibinfo{author}{\bibfnamefont{A.}~\bibnamefont{Vishwanath}},
  \bibinfo{journal}{Rev. Mod. Phys.} \textbf{\bibinfo{volume}{90}},
  \bibinfo{pages}{015001} (\bibinfo{year}{2018}).

\bibitem[{\citenamefont{Burkov et~al.}(2011)\citenamefont{Burkov, Hook, and
  Balents}}]{Burkov2011Topological}
\bibinfo{author}{\bibfnamefont{A.~A.} \bibnamefont{Burkov}},
  \bibinfo{author}{\bibfnamefont{M.~D.} \bibnamefont{Hook}}, \bibnamefont{and}
  \bibinfo{author}{\bibfnamefont{L.}~\bibnamefont{Balents}},
  \bibinfo{journal}{Phys. Rev. B} \textbf{\bibinfo{volume}{84}},
  \bibinfo{pages}{235126} (\bibinfo{year}{2011}).

\bibitem[{\citenamefont{Fang et~al.}(2016)\citenamefont{Fang, Weng, Dai, and
  Fang}}]{Fang2016Topological}
\bibinfo{author}{\bibfnamefont{C.}~\bibnamefont{Fang}},
  \bibinfo{author}{\bibfnamefont{H.}~\bibnamefont{Weng}},
  \bibinfo{author}{\bibfnamefont{X.}~\bibnamefont{Dai}}, \bibnamefont{and}
  \bibinfo{author}{\bibfnamefont{Z.}~\bibnamefont{Fang}},
  \bibinfo{journal}{Chin. Phys. B} \textbf{\bibinfo{volume}{25}},
  \bibinfo{pages}{117106} (\bibinfo{year}{2016}).

\bibitem[{\citenamefont{Yu et~al.}(2017)\citenamefont{Yu, Fang, Dai, and
  Weng}}]{Yu2017Topological}
\bibinfo{author}{\bibfnamefont{R.}~\bibnamefont{Yu}},
  \bibinfo{author}{\bibfnamefont{Z.}~\bibnamefont{Fang}},
  \bibinfo{author}{\bibfnamefont{X.}~\bibnamefont{Dai}}, \bibnamefont{and}
  \bibinfo{author}{\bibfnamefont{H.}~\bibnamefont{Weng}},
  \bibinfo{journal}{Front. Phys.} \textbf{\bibinfo{volume}{12}},
  \bibinfo{pages}{127202} (\bibinfo{year}{2017}).

\bibitem[{\citenamefont{Yang et~al.}(2018)\citenamefont{Yang, Yang, Derunova,
  Parkin, Yan, and Ali}}]{Yang2018Symmetry}
\bibinfo{author}{\bibfnamefont{S.-Y.} \bibnamefont{Yang}},
  \bibinfo{author}{\bibfnamefont{H.}~\bibnamefont{Yang}},
  \bibinfo{author}{\bibfnamefont{E.}~\bibnamefont{Derunova}},
  \bibinfo{author}{\bibfnamefont{S.~S.~P.} \bibnamefont{Parkin}},
  \bibinfo{author}{\bibfnamefont{B.}~\bibnamefont{Yan}}, \bibnamefont{and}
  \bibinfo{author}{\bibfnamefont{M.~N.} \bibnamefont{Ali}},
  \bibinfo{journal}{Adv. Phys. X} \textbf{\bibinfo{volume}{3}},
  \bibinfo{pages}{1414631} (\bibinfo{year}{2018}).

\bibitem[{\citenamefont{Song et~al.}(2020)\citenamefont{Song, Wang, Li, Liu,
  Lu, Liu, Li, Wen, Yin, Liu et~al.}}]{Song2020Photoemission}
\bibinfo{author}{\bibfnamefont{Y.~K.} \bibnamefont{Song}},
  \bibinfo{author}{\bibfnamefont{G.~W.} \bibnamefont{Wang}},
  \bibinfo{author}{\bibfnamefont{S.~C.} \bibnamefont{Li}},
  \bibinfo{author}{\bibfnamefont{W.~L.} \bibnamefont{Liu}},
  \bibinfo{author}{\bibfnamefont{X.~L.} \bibnamefont{Lu}},
  \bibinfo{author}{\bibfnamefont{Z.~T.} \bibnamefont{Liu}},
  \bibinfo{author}{\bibfnamefont{Z.~J.} \bibnamefont{Li}},
  \bibinfo{author}{\bibfnamefont{J.~S.} \bibnamefont{Wen}},
  \bibinfo{author}{\bibfnamefont{Z.~P.} \bibnamefont{Yin}},
  \bibinfo{author}{\bibfnamefont{Z.~H.} \bibnamefont{Liu}},
  \bibnamefont{et~al.}, \bibinfo{journal}{Phys. Rev. Lett.}
  \textbf{\bibinfo{volume}{124}}, \bibinfo{pages}{056402}
  (\bibinfo{year}{2020}).

\bibitem[{\citenamefont{Liu et~al.}(2018)\citenamefont{Liu, Lou, Guo, Wang,
  Sun, Li, Thirupathaiah, Fedorov, Shen, Liu et~al.}}]{Liu2018Experimental}
\bibinfo{author}{\bibfnamefont{Z.}~\bibnamefont{Liu}},
  \bibinfo{author}{\bibfnamefont{R.}~\bibnamefont{Lou}},
  \bibinfo{author}{\bibfnamefont{P.}~\bibnamefont{Guo}},
  \bibinfo{author}{\bibfnamefont{Q.}~\bibnamefont{Wang}},
  \bibinfo{author}{\bibfnamefont{S.}~\bibnamefont{Sun}},
  \bibinfo{author}{\bibfnamefont{C.}~\bibnamefont{Li}},
  \bibinfo{author}{\bibfnamefont{S.}~\bibnamefont{Thirupathaiah}},
  \bibinfo{author}{\bibfnamefont{A.}~\bibnamefont{Fedorov}},
  \bibinfo{author}{\bibfnamefont{D.}~\bibnamefont{Shen}},
  \bibinfo{author}{\bibfnamefont{K.}~\bibnamefont{Liu}}, \bibnamefont{et~al.},
  \bibinfo{journal}{Phys. Rev. X} \textbf{\bibinfo{volume}{8}},
  \bibinfo{pages}{031044} (\bibinfo{year}{2018}).

\bibitem[{\citenamefont{Bzdusek et~al.}(2016)\citenamefont{Bzdusek, Wu, Ruegg,
  Sigrist, and Soluyanov}}]{Bzdusek2016Nodal}
\bibinfo{author}{\bibfnamefont{T.}~\bibnamefont{Bzdusek}},
  \bibinfo{author}{\bibfnamefont{Q.}~\bibnamefont{Wu}},
  \bibinfo{author}{\bibfnamefont{A.}~\bibnamefont{Ruegg}},
  \bibinfo{author}{\bibfnamefont{M.}~\bibnamefont{Sigrist}}, \bibnamefont{and}
  \bibinfo{author}{\bibfnamefont{A.~A.} \bibnamefont{Soluyanov}},
  \bibinfo{journal}{Nature} \textbf{\bibinfo{volume}{538}}, \bibinfo{pages}{75}
  (\bibinfo{year}{2016}).

\bibitem[{\citenamefont{Yan et~al.}(2018)\citenamefont{Yan, Liu, Yan, Liu,
  Chen, Wang, and Lu}}]{Yan2018Experimental}
\bibinfo{author}{\bibfnamefont{Q.}~\bibnamefont{Yan}},
  \bibinfo{author}{\bibfnamefont{R.}~\bibnamefont{Liu}},
  \bibinfo{author}{\bibfnamefont{Z.}~\bibnamefont{Yan}},
  \bibinfo{author}{\bibfnamefont{B.}~\bibnamefont{Liu}},
  \bibinfo{author}{\bibfnamefont{H.}~\bibnamefont{Chen}},
  \bibinfo{author}{\bibfnamefont{Z.}~\bibnamefont{Wang}}, \bibnamefont{and}
  \bibinfo{author}{\bibfnamefont{L.}~\bibnamefont{Lu}}, \bibinfo{journal}{Nat.
  Phys.} \textbf{\bibinfo{volume}{14}}, \bibinfo{pages}{461}
  (\bibinfo{year}{2018}).

\bibitem[{\citenamefont{Wu et~al.}(2018)\citenamefont{Wu, Liu, Li, Zhong, Yu,
  Sheng, Zhao, and Yang}}]{Wu2018Nodal}
\bibinfo{author}{\bibfnamefont{W.}~\bibnamefont{Wu}},
  \bibinfo{author}{\bibfnamefont{Y.}~\bibnamefont{Liu}},
  \bibinfo{author}{\bibfnamefont{S.}~\bibnamefont{Li}},
  \bibinfo{author}{\bibfnamefont{C.}~\bibnamefont{Zhong}},
  \bibinfo{author}{\bibfnamefont{Z.-M.} \bibnamefont{Yu}},
  \bibinfo{author}{\bibfnamefont{X.-L.} \bibnamefont{Sheng}},
  \bibinfo{author}{\bibfnamefont{Y.~X.} \bibnamefont{Zhao}}, \bibnamefont{and}
  \bibinfo{author}{\bibfnamefont{S.~A.} \bibnamefont{Yang}},
  \bibinfo{journal}{Phys. Rev. B} \textbf{\bibinfo{volume}{97}},
  \bibinfo{pages}{115125} (\bibinfo{year}{2018}).

\bibitem[{\citenamefont{Fu et~al.}(2019)\citenamefont{Fu, Yi, Zhang, Caputo,
  Ma, Gao, Lv, Kong, Huang, Richard et~al.}}]{Fu2019Dirac}
\bibinfo{author}{\bibfnamefont{B.~B.} \bibnamefont{Fu}},
  \bibinfo{author}{\bibfnamefont{C.~J.} \bibnamefont{Yi}},
  \bibinfo{author}{\bibfnamefont{T.~T.} \bibnamefont{Zhang}},
  \bibinfo{author}{\bibfnamefont{M.}~\bibnamefont{Caputo}},
  \bibinfo{author}{\bibfnamefont{J.~Z.} \bibnamefont{Ma}},
  \bibinfo{author}{\bibfnamefont{X.}~\bibnamefont{Gao}},
  \bibinfo{author}{\bibfnamefont{B.~Q.} \bibnamefont{Lv}},
  \bibinfo{author}{\bibfnamefont{L.~Y.} \bibnamefont{Kong}},
  \bibinfo{author}{\bibfnamefont{Y.~B.} \bibnamefont{Huang}},
  \bibinfo{author}{\bibfnamefont{P.}~\bibnamefont{Richard}},
  \bibnamefont{et~al.}, \bibinfo{journal}{Sci. Adv.}
  \textbf{\bibinfo{volume}{5}}, \bibinfo{pages}{eaau6459}
  (\bibinfo{year}{2019}).

\bibitem[{\citenamefont{Song et~al.}(2022)\citenamefont{Song, Jin, Song, Rong,
  Zhu, Liang, Cui, Sun, Zhao, Shi et~al.}}]{Song2022Spectroscopic}
\bibinfo{author}{\bibfnamefont{C.}~\bibnamefont{Song}},
  \bibinfo{author}{\bibfnamefont{L.}~\bibnamefont{Jin}},
  \bibinfo{author}{\bibfnamefont{P.}~\bibnamefont{Song}},
  \bibinfo{author}{\bibfnamefont{H.}~\bibnamefont{Rong}},
  \bibinfo{author}{\bibfnamefont{W.}~\bibnamefont{Zhu}},
  \bibinfo{author}{\bibfnamefont{B.}~\bibnamefont{Liang}},
  \bibinfo{author}{\bibfnamefont{S.}~\bibnamefont{Cui}},
  \bibinfo{author}{\bibfnamefont{Z.}~\bibnamefont{Sun}},
  \bibinfo{author}{\bibfnamefont{L.}~\bibnamefont{Zhao}},
  \bibinfo{author}{\bibfnamefont{Y.}~\bibnamefont{Shi}}, \bibnamefont{et~al.},
  \bibinfo{journal}{Phys. Rev. B} \textbf{\bibinfo{volume}{105}},
  \bibinfo{pages}{L161104} (\bibinfo{year}{2022}).

\bibitem[{\citenamefont{Xu et~al.}(2015{\natexlab{b}})\citenamefont{Xu, Liu,
  Kushwaha, Sankar, Krizan, Belopolski, Neupane, Bian, Alidoust, Chang
  et~al.}}]{Xu2015Observation}
\bibinfo{author}{\bibfnamefont{S.~Y.} \bibnamefont{Xu}},
  \bibinfo{author}{\bibfnamefont{C.}~\bibnamefont{Liu}},
  \bibinfo{author}{\bibfnamefont{S.~K.} \bibnamefont{Kushwaha}},
  \bibinfo{author}{\bibfnamefont{R.}~\bibnamefont{Sankar}},
  \bibinfo{author}{\bibfnamefont{J.~W.} \bibnamefont{Krizan}},
  \bibinfo{author}{\bibfnamefont{I.}~\bibnamefont{Belopolski}},
  \bibinfo{author}{\bibfnamefont{M.}~\bibnamefont{Neupane}},
  \bibinfo{author}{\bibfnamefont{G.}~\bibnamefont{Bian}},
  \bibinfo{author}{\bibfnamefont{N.}~\bibnamefont{Alidoust}},
  \bibinfo{author}{\bibfnamefont{T.~R.} \bibnamefont{Chang}},
  \bibnamefont{et~al.}, \bibinfo{journal}{Science}
  \textbf{\bibinfo{volume}{347}}, \bibinfo{pages}{294}
  (\bibinfo{year}{2015}{\natexlab{b}}).

\bibitem[{\citenamefont{Rao et~al.}(2019)\citenamefont{Rao, Li, Zhang, Tian,
  Li, Fu, Tang, Wang, Li, Fan et~al.}}]{Rao2019Observation}
\bibinfo{author}{\bibfnamefont{Z.}~\bibnamefont{Rao}},
  \bibinfo{author}{\bibfnamefont{H.}~\bibnamefont{Li}},
  \bibinfo{author}{\bibfnamefont{T.}~\bibnamefont{Zhang}},
  \bibinfo{author}{\bibfnamefont{S.}~\bibnamefont{Tian}},
  \bibinfo{author}{\bibfnamefont{C.}~\bibnamefont{Li}},
  \bibinfo{author}{\bibfnamefont{B.}~\bibnamefont{Fu}},
  \bibinfo{author}{\bibfnamefont{C.}~\bibnamefont{Tang}},
  \bibinfo{author}{\bibfnamefont{L.}~\bibnamefont{Wang}},
  \bibinfo{author}{\bibfnamefont{Z.}~\bibnamefont{Li}},
  \bibinfo{author}{\bibfnamefont{W.}~\bibnamefont{Fan}}, \bibnamefont{et~al.},
  \bibinfo{journal}{Nature} \textbf{\bibinfo{volume}{567}},
  \bibinfo{pages}{496} (\bibinfo{year}{2019}).

\bibitem[{\citenamefont{Bian et~al.}(2016)\citenamefont{Bian, Chang, Sankar,
  Xu, Zheng, Neupert, Chiu, Huang, Chang, Belopolski
  et~al.}}]{Bian2016Topological}
\bibinfo{author}{\bibfnamefont{G.}~\bibnamefont{Bian}},
  \bibinfo{author}{\bibfnamefont{T.~R.} \bibnamefont{Chang}},
  \bibinfo{author}{\bibfnamefont{R.}~\bibnamefont{Sankar}},
  \bibinfo{author}{\bibfnamefont{S.~Y.} \bibnamefont{Xu}},
  \bibinfo{author}{\bibfnamefont{H.}~\bibnamefont{Zheng}},
  \bibinfo{author}{\bibfnamefont{T.}~\bibnamefont{Neupert}},
  \bibinfo{author}{\bibfnamefont{C.~K.} \bibnamefont{Chiu}},
  \bibinfo{author}{\bibfnamefont{S.~M.} \bibnamefont{Huang}},
  \bibinfo{author}{\bibfnamefont{G.}~\bibnamefont{Chang}},
  \bibinfo{author}{\bibfnamefont{I.}~\bibnamefont{Belopolski}},
  \bibnamefont{et~al.}, \bibinfo{journal}{Nat. Commun.}
  \textbf{\bibinfo{volume}{7}}, \bibinfo{pages}{10556} (\bibinfo{year}{2016}).

\bibitem[{\citenamefont{Schoop et~al.}(2016)\citenamefont{Schoop, Ali,
  Strasser, Topp, Varykhalov, Marchenko, Duppel, Parkin, Lotsch, and
  Ast}}]{Schoop2016Dirac}
\bibinfo{author}{\bibfnamefont{L.~M.} \bibnamefont{Schoop}},
  \bibinfo{author}{\bibfnamefont{M.~N.} \bibnamefont{Ali}},
  \bibinfo{author}{\bibfnamefont{C.}~\bibnamefont{Strasser}},
  \bibinfo{author}{\bibfnamefont{A.}~\bibnamefont{Topp}},
  \bibinfo{author}{\bibfnamefont{A.}~\bibnamefont{Varykhalov}},
  \bibinfo{author}{\bibfnamefont{D.}~\bibnamefont{Marchenko}},
  \bibinfo{author}{\bibfnamefont{V.}~\bibnamefont{Duppel}},
  \bibinfo{author}{\bibfnamefont{S.~S.} \bibnamefont{Parkin}},
  \bibinfo{author}{\bibfnamefont{B.~V.} \bibnamefont{Lotsch}},
  \bibnamefont{and} \bibinfo{author}{\bibfnamefont{C.~R.} \bibnamefont{Ast}},
  \bibinfo{journal}{Nat. Commun.} \textbf{\bibinfo{volume}{7}},
  \bibinfo{pages}{11696} (\bibinfo{year}{2016}).

\bibitem[{\citenamefont{Ali et~al.}(2014)\citenamefont{Ali, Xiong, Flynn, Tao,
  Gibson, Schoop, Liang, Haldolaarachchige, Hirschberger, Ong
  et~al.}}]{Ali2014Large}
\bibinfo{author}{\bibfnamefont{M.~N.} \bibnamefont{Ali}},
  \bibinfo{author}{\bibfnamefont{J.}~\bibnamefont{Xiong}},
  \bibinfo{author}{\bibfnamefont{S.}~\bibnamefont{Flynn}},
  \bibinfo{author}{\bibfnamefont{J.}~\bibnamefont{Tao}},
  \bibinfo{author}{\bibfnamefont{Q.~D.} \bibnamefont{Gibson}},
  \bibinfo{author}{\bibfnamefont{L.~M.} \bibnamefont{Schoop}},
  \bibinfo{author}{\bibfnamefont{T.}~\bibnamefont{Liang}},
  \bibinfo{author}{\bibfnamefont{N.}~\bibnamefont{Haldolaarachchige}},
  \bibinfo{author}{\bibfnamefont{M.}~\bibnamefont{Hirschberger}},
  \bibinfo{author}{\bibfnamefont{N.~P.} \bibnamefont{Ong}},
  \bibnamefont{et~al.}, \bibinfo{journal}{Nature}
  \textbf{\bibinfo{volume}{514}}, \bibinfo{pages}{205} (\bibinfo{year}{2014}).

\bibitem[{\citenamefont{Huang et~al.}(2015)\citenamefont{Huang, Zhao, Long,
  Wang, Chen, Yang, Liang, Xue, Weng, Fang et~al.}}]{Huang2015Observation}
\bibinfo{author}{\bibfnamefont{X.}~\bibnamefont{Huang}},
  \bibinfo{author}{\bibfnamefont{L.}~\bibnamefont{Zhao}},
  \bibinfo{author}{\bibfnamefont{Y.}~\bibnamefont{Long}},
  \bibinfo{author}{\bibfnamefont{P.}~\bibnamefont{Wang}},
  \bibinfo{author}{\bibfnamefont{D.}~\bibnamefont{Chen}},
  \bibinfo{author}{\bibfnamefont{Z.}~\bibnamefont{Yang}},
  \bibinfo{author}{\bibfnamefont{H.}~\bibnamefont{Liang}},
  \bibinfo{author}{\bibfnamefont{M.}~\bibnamefont{Xue}},
  \bibinfo{author}{\bibfnamefont{H.}~\bibnamefont{Weng}},
  \bibinfo{author}{\bibfnamefont{Z.}~\bibnamefont{Fang}}, \bibnamefont{et~al.},
  \bibinfo{journal}{Phys. Rev. X} \textbf{\bibinfo{volume}{5}},
  \bibinfo{pages}{031023} (\bibinfo{year}{2015}).

\bibitem[{\citenamefont{Li et~al.}(2016)\citenamefont{Li, He, Lu, Zhang, Liu,
  Ma, Fan, Shen, and Wang}}]{Li2016Negative}
\bibinfo{author}{\bibfnamefont{H.}~\bibnamefont{Li}},
  \bibinfo{author}{\bibfnamefont{H.}~\bibnamefont{He}},
  \bibinfo{author}{\bibfnamefont{H.~Z.} \bibnamefont{Lu}},
  \bibinfo{author}{\bibfnamefont{H.}~\bibnamefont{Zhang}},
  \bibinfo{author}{\bibfnamefont{H.}~\bibnamefont{Liu}},
  \bibinfo{author}{\bibfnamefont{R.}~\bibnamefont{Ma}},
  \bibinfo{author}{\bibfnamefont{Z.}~\bibnamefont{Fan}},
  \bibinfo{author}{\bibfnamefont{S.~Q.} \bibnamefont{Shen}}, \bibnamefont{and}
  \bibinfo{author}{\bibfnamefont{J.}~\bibnamefont{Wang}},
  \bibinfo{journal}{Nat. Commun.} \textbf{\bibinfo{volume}{7}},
  \bibinfo{pages}{10301} (\bibinfo{year}{2016}).

\bibitem[{\citenamefont{Fukushima et~al.}(2008)\citenamefont{Fukushima,
  Kharzeev, and Warringa}}]{Fukushima2008Chiral}
\bibinfo{author}{\bibfnamefont{K.}~\bibnamefont{Fukushima}},
  \bibinfo{author}{\bibfnamefont{D.~E.} \bibnamefont{Kharzeev}},
  \bibnamefont{and} \bibinfo{author}{\bibfnamefont{H.~J.}
  \bibnamefont{Warringa}}, \bibinfo{journal}{Phys. Rev. D}
  \textbf{\bibinfo{volume}{78}}, \bibinfo{pages}{074033}
  (\bibinfo{year}{2008}).

\bibitem[{\citenamefont{Schröter et~al.}(2019)\citenamefont{Schröter, Pei,
  Vergniory, Sun, Manna, de~Juan, Krieger, Süss, Schmidt, Dudin
  et~al.}}]{Schroter2019Chiral}
\bibinfo{author}{\bibfnamefont{N.~B.~M.} \bibnamefont{Schröter}},
  \bibinfo{author}{\bibfnamefont{D.}~\bibnamefont{Pei}},
  \bibinfo{author}{\bibfnamefont{M.~G.} \bibnamefont{Vergniory}},
  \bibinfo{author}{\bibfnamefont{Y.}~\bibnamefont{Sun}},
  \bibinfo{author}{\bibfnamefont{K.}~\bibnamefont{Manna}},
  \bibinfo{author}{\bibfnamefont{F.}~\bibnamefont{de~Juan}},
  \bibinfo{author}{\bibfnamefont{J.~A.} \bibnamefont{Krieger}},
  \bibinfo{author}{\bibfnamefont{V.}~\bibnamefont{Süss}},
  \bibinfo{author}{\bibfnamefont{M.}~\bibnamefont{Schmidt}},
  \bibinfo{author}{\bibfnamefont{P.}~\bibnamefont{Dudin}},
  \bibnamefont{et~al.}, \bibinfo{journal}{Nat. Phys.}
  \textbf{\bibinfo{volume}{15}}, \bibinfo{pages}{759} (\bibinfo{year}{2019}).

\bibitem[{\citenamefont{Bradlyn et~al.}(2016)\citenamefont{Bradlyn, Cano, Wang,
  Vergniory, Felser, Cava, and Bernevig}}]{Bradlyn2016Beyond}
\bibinfo{author}{\bibfnamefont{B.}~\bibnamefont{Bradlyn}},
  \bibinfo{author}{\bibfnamefont{J.}~\bibnamefont{Cano}},
  \bibinfo{author}{\bibfnamefont{Z.}~\bibnamefont{Wang}},
  \bibinfo{author}{\bibfnamefont{M.~G.} \bibnamefont{Vergniory}},
  \bibinfo{author}{\bibfnamefont{C.}~\bibnamefont{Felser}},
  \bibinfo{author}{\bibfnamefont{R.~J.} \bibnamefont{Cava}}, \bibnamefont{and}
  \bibinfo{author}{\bibfnamefont{B.~A.} \bibnamefont{Bernevig}},
  \bibinfo{journal}{Science} \textbf{\bibinfo{volume}{353}},
  \bibinfo{pages}{aaf5037} (\bibinfo{year}{2016}).

\bibitem[{\citenamefont{Guo et~al.}(2021)\citenamefont{Guo, Wei, Liu, Liu, and
  Lu}}]{Guo2021Eightfold}
\bibinfo{author}{\bibfnamefont{P.-J.} \bibnamefont{Guo}},
  \bibinfo{author}{\bibfnamefont{Y.-W.} \bibnamefont{Wei}},
  \bibinfo{author}{\bibfnamefont{K.}~\bibnamefont{Liu}},
  \bibinfo{author}{\bibfnamefont{Z.-X.} \bibnamefont{Liu}}, \bibnamefont{and}
  \bibinfo{author}{\bibfnamefont{Z.-Y.} \bibnamefont{Lu}},
  \bibinfo{journal}{Phys. Rev. Lett.} \textbf{\bibinfo{volume}{127}},
  \bibinfo{pages}{176401} (\bibinfo{year}{2021}).

\bibitem[{\citenamefont{Sun et~al.}(2017)\citenamefont{Sun, Zhang, and
  Chang}}]{Sun2017Coexistence}
\bibinfo{author}{\bibfnamefont{J.-P.} \bibnamefont{Sun}},
  \bibinfo{author}{\bibfnamefont{D.}~\bibnamefont{Zhang}}, \bibnamefont{and}
  \bibinfo{author}{\bibfnamefont{K.}~\bibnamefont{Chang}},
  \bibinfo{journal}{Phys. Rev. B} \textbf{\bibinfo{volume}{96}},
  \bibinfo{pages}{045121} (\bibinfo{year}{2017}).

\bibitem[{\citenamefont{Tang et~al.}(2017)\citenamefont{Tang, Zhou, and
  Zhang}}]{Tang2017Multiple}
\bibinfo{author}{\bibfnamefont{P.}~\bibnamefont{Tang}},
  \bibinfo{author}{\bibfnamefont{Q.}~\bibnamefont{Zhou}}, \bibnamefont{and}
  \bibinfo{author}{\bibfnamefont{S.-C.} \bibnamefont{Zhang}},
  \bibinfo{journal}{Phys. Rev. Lett.} \textbf{\bibinfo{volume}{119}},
  \bibinfo{pages}{206402} (\bibinfo{year}{2017}).

\bibitem[{\citenamefont{Wu et~al.}(2021)\citenamefont{Wu, Tang, and
  Wan}}]{Wu2021Symmetry}
\bibinfo{author}{\bibfnamefont{L.}~\bibnamefont{Wu}},
  \bibinfo{author}{\bibfnamefont{F.}~\bibnamefont{Tang}}, \bibnamefont{and}
  \bibinfo{author}{\bibfnamefont{X.}~\bibnamefont{Wan}},
  \bibinfo{journal}{Phys. Rev. B} \textbf{\bibinfo{volume}{104}},
  \bibinfo{pages}{045107} (\bibinfo{year}{2021}).

\bibitem[{\citenamefont{Xia and Li}(2017)}]{Xia2017Triply}
\bibinfo{author}{\bibfnamefont{Y.}~\bibnamefont{Xia}} \bibnamefont{and}
  \bibinfo{author}{\bibfnamefont{G.}~\bibnamefont{Li}}, \bibinfo{journal}{Phys.
  Rev. B} \textbf{\bibinfo{volume}{96}}, \bibinfo{pages}{241204}
  (\bibinfo{year}{2017}).

\bibitem[{\citenamefont{Zhang et~al.}(2017)\citenamefont{Zhang, Yu, Sheng,
  Yang, and Yang}}]{Zhang2017Coexistence}
\bibinfo{author}{\bibfnamefont{X.}~\bibnamefont{Zhang}},
  \bibinfo{author}{\bibfnamefont{Z.-M.} \bibnamefont{Yu}},
  \bibinfo{author}{\bibfnamefont{X.-L.} \bibnamefont{Sheng}},
  \bibinfo{author}{\bibfnamefont{H.~Y.} \bibnamefont{Yang}}, \bibnamefont{and}
  \bibinfo{author}{\bibfnamefont{S.~A.} \bibnamefont{Yang}},
  \bibinfo{journal}{Phys. Rev. B} \textbf{\bibinfo{volume}{95}},
  \bibinfo{pages}{235116} (\bibinfo{year}{2017}).

\bibitem[{\citenamefont{Lv et~al.}(2017)\citenamefont{Lv, Feng, Xu, Gao, Ma,
  Kong, Richard, Huang, Strocov, Fang et~al.}}]{Lv2017Observation}
\bibinfo{author}{\bibfnamefont{B.~Q.} \bibnamefont{Lv}},
  \bibinfo{author}{\bibfnamefont{Z.~L.} \bibnamefont{Feng}},
  \bibinfo{author}{\bibfnamefont{Q.~N.} \bibnamefont{Xu}},
  \bibinfo{author}{\bibfnamefont{X.}~\bibnamefont{Gao}},
  \bibinfo{author}{\bibfnamefont{J.~Z.} \bibnamefont{Ma}},
  \bibinfo{author}{\bibfnamefont{L.~Y.} \bibnamefont{Kong}},
  \bibinfo{author}{\bibfnamefont{P.}~\bibnamefont{Richard}},
  \bibinfo{author}{\bibfnamefont{Y.~B.} \bibnamefont{Huang}},
  \bibinfo{author}{\bibfnamefont{V.~N.} \bibnamefont{Strocov}},
  \bibinfo{author}{\bibfnamefont{C.}~\bibnamefont{Fang}}, \bibnamefont{et~al.},
  \bibinfo{journal}{Nature} \textbf{\bibinfo{volume}{546}},
  \bibinfo{pages}{627} (\bibinfo{year}{2017}).

\bibitem[{\citenamefont{Ma et~al.}(2018)\citenamefont{Ma, He, Xu, Lv, Chen,
  Zhu, Zhang, Kong, Gao, Rong et~al.}}]{Ma2018Threecomponent}
\bibinfo{author}{\bibfnamefont{J.~Z.} \bibnamefont{Ma}},
  \bibinfo{author}{\bibfnamefont{J.~B.} \bibnamefont{He}},
  \bibinfo{author}{\bibfnamefont{Y.~F.} \bibnamefont{Xu}},
  \bibinfo{author}{\bibfnamefont{B.~Q.} \bibnamefont{Lv}},
  \bibinfo{author}{\bibfnamefont{D.}~\bibnamefont{Chen}},
  \bibinfo{author}{\bibfnamefont{W.~L.} \bibnamefont{Zhu}},
  \bibinfo{author}{\bibfnamefont{S.}~\bibnamefont{Zhang}},
  \bibinfo{author}{\bibfnamefont{L.~Y.} \bibnamefont{Kong}},
  \bibinfo{author}{\bibfnamefont{X.}~\bibnamefont{Gao}},
  \bibinfo{author}{\bibfnamefont{L.~Y.} \bibnamefont{Rong}},
  \bibnamefont{et~al.}, \bibinfo{journal}{Nat. Phys.}
  \textbf{\bibinfo{volume}{14}}, \bibinfo{pages}{349} (\bibinfo{year}{2018}).

\bibitem[{\citenamefont{Sun et~al.}(2020)\citenamefont{Sun, Hua, Liu, Liu, Ye,
  Qiao, Liu, Liu, Guo, Lu et~al.}}]{Sun2020Direct}
\bibinfo{author}{\bibfnamefont{Z.~P.} \bibnamefont{Sun}},
  \bibinfo{author}{\bibfnamefont{C.~Q.} \bibnamefont{Hua}},
  \bibinfo{author}{\bibfnamefont{X.~L.} \bibnamefont{Liu}},
  \bibinfo{author}{\bibfnamefont{Z.~T.} \bibnamefont{Liu}},
  \bibinfo{author}{\bibfnamefont{M.}~\bibnamefont{Ye}},
  \bibinfo{author}{\bibfnamefont{S.}~\bibnamefont{Qiao}},
  \bibinfo{author}{\bibfnamefont{Z.~H.} \bibnamefont{Liu}},
  \bibinfo{author}{\bibfnamefont{J.~S.} \bibnamefont{Liu}},
  \bibinfo{author}{\bibfnamefont{Y.~F.} \bibnamefont{Guo}},
  \bibinfo{author}{\bibfnamefont{Y.~H.} \bibnamefont{Lu}},
  \bibnamefont{et~al.}, \bibinfo{journal}{Phys. Rev. B}
  \textbf{\bibinfo{volume}{101}}, \bibinfo{pages}{155114}
  (\bibinfo{year}{2020}).

\bibitem[{\citenamefont{Schoop et~al.}(2018)\citenamefont{Schoop, Topp,
  Lippmann, Orlandi, Muchler, Vergniory, Sun, Rost, Duppel, Krivenkov
  et~al.}}]{Schoop2018Tunable}
\bibinfo{author}{\bibfnamefont{L.~M.} \bibnamefont{Schoop}},
  \bibinfo{author}{\bibfnamefont{A.}~\bibnamefont{Topp}},
  \bibinfo{author}{\bibfnamefont{J.}~\bibnamefont{Lippmann}},
  \bibinfo{author}{\bibfnamefont{F.}~\bibnamefont{Orlandi}},
  \bibinfo{author}{\bibfnamefont{L.}~\bibnamefont{Muchler}},
  \bibinfo{author}{\bibfnamefont{M.~G.} \bibnamefont{Vergniory}},
  \bibinfo{author}{\bibfnamefont{Y.}~\bibnamefont{Sun}},
  \bibinfo{author}{\bibfnamefont{A.~W.} \bibnamefont{Rost}},
  \bibinfo{author}{\bibfnamefont{V.}~\bibnamefont{Duppel}},
  \bibinfo{author}{\bibfnamefont{M.}~\bibnamefont{Krivenkov}},
  \bibnamefont{et~al.}, \bibinfo{journal}{Sci. Adv.}
  \textbf{\bibinfo{volume}{4}}, \bibinfo{pages}{eaar2317}
  (\bibinfo{year}{2018}).

\bibitem[{\citenamefont{Berry et~al.}(2020)\citenamefont{Berry, Pressley,
  Phelan, Tran, and McQueen}}]{Berry2020Laser}
\bibinfo{author}{\bibfnamefont{T.}~\bibnamefont{Berry}},
  \bibinfo{author}{\bibfnamefont{L.~A.} \bibnamefont{Pressley}},
  \bibinfo{author}{\bibfnamefont{W.~A.} \bibnamefont{Phelan}},
  \bibinfo{author}{\bibfnamefont{T.~T.} \bibnamefont{Tran}}, \bibnamefont{and}
  \bibinfo{author}{\bibfnamefont{T.~M.} \bibnamefont{McQueen}},
  \bibinfo{journal}{Chem. Mater.} \textbf{\bibinfo{volume}{32}},
  \bibinfo{pages}{5827} (\bibinfo{year}{2020}).

\bibitem[{\citenamefont{{Palacio} et~al.}(2022)\citenamefont{{Palacio},
  {Obando-Guevara}, {Chen}, {Nair}, {Gonz{\'a}lez Barrio}, {Papalazarou}, {Le
  F{\`e}vre}, {Luccas}, {Suderow}, {Canfield} et~al.}}]{2022arXiv220203161P}
\bibinfo{author}{\bibfnamefont{I.}~\bibnamefont{{Palacio}}},
  \bibinfo{author}{\bibfnamefont{J.}~\bibnamefont{{Obando-Guevara}}},
  \bibinfo{author}{\bibfnamefont{L.}~\bibnamefont{{Chen}}},
  \bibinfo{author}{\bibfnamefont{M.~N.} \bibnamefont{{Nair}}},
  \bibinfo{author}{\bibfnamefont{M.~A.} \bibnamefont{{Gonz{\'a}lez Barrio}}},
  \bibinfo{author}{\bibfnamefont{E.}~\bibnamefont{{Papalazarou}}},
  \bibinfo{author}{\bibfnamefont{P.}~\bibnamefont{{Le F{\`e}vre}}},
  \bibinfo{author}{\bibfnamefont{R.~F.} \bibnamefont{{Luccas}}},
  \bibinfo{author}{\bibfnamefont{H.}~\bibnamefont{{Suderow}}},
  \bibinfo{author}{\bibfnamefont{P.}~\bibnamefont{{Canfield}}},
  \bibnamefont{et~al.}, \bibinfo{journal}{arXiv:2202.03161}
  (\bibinfo{year}{2022}).

\bibitem[{\citenamefont{{Bud'ko} et~al.}(2022)\citenamefont{{Bud'ko}, {Huyan},
  {Herrera-Siklody}, and {Canfield}}}]{2022arXiv220804997B}
\bibinfo{author}{\bibfnamefont{S.~L.} \bibnamefont{{Bud'ko}}},
  \bibinfo{author}{\bibfnamefont{S.}~\bibnamefont{{Huyan}}},
  \bibinfo{author}{\bibfnamefont{P.}~\bibnamefont{{Herrera-Siklody}}},
  \bibnamefont{and} \bibinfo{author}{\bibfnamefont{P.~C.}
  \bibnamefont{{Canfield}}}, \bibinfo{journal}{arXiv:2208.04997}
  (\bibinfo{year}{2022}).

\bibitem[{\citenamefont{Guo et~al.}(2011)\citenamefont{Guo, Young, Adams, Wu,
  Chan, McCandless, and DiTusa}}]{Guo2011Dimensional}
\bibinfo{author}{\bibfnamefont{S.}~\bibnamefont{Guo}},
  \bibinfo{author}{\bibfnamefont{D.~P.} \bibnamefont{Young}},
  \bibinfo{author}{\bibfnamefont{P.~W.} \bibnamefont{Adams}},
  \bibinfo{author}{\bibfnamefont{X.~S.} \bibnamefont{Wu}},
  \bibinfo{author}{\bibfnamefont{J.~Y.} \bibnamefont{Chan}},
  \bibinfo{author}{\bibfnamefont{G.~T.} \bibnamefont{McCandless}},
  \bibnamefont{and} \bibinfo{author}{\bibfnamefont{J.~F.}
  \bibnamefont{DiTusa}}, \bibinfo{journal}{Phys. Rev. B}
  \textbf{\bibinfo{volume}{83}}, \bibinfo{pages}{174520}
  (\bibinfo{year}{2011}).

\bibitem[{\citenamefont{Galvis et~al.}(2013)\citenamefont{Galvis, Suderow,
  Vieira, Bud'ko, and Canfield}}]{Galvis2013Scanning}
\bibinfo{author}{\bibfnamefont{J.~A.} \bibnamefont{Galvis}},
  \bibinfo{author}{\bibfnamefont{H.}~\bibnamefont{Suderow}},
  \bibinfo{author}{\bibfnamefont{S.}~\bibnamefont{Vieira}},
  \bibinfo{author}{\bibfnamefont{S.~L.} \bibnamefont{Bud'ko}},
  \bibnamefont{and} \bibinfo{author}{\bibfnamefont{P.~C.}
  \bibnamefont{Canfield}}, \bibinfo{journal}{Phys. Rev. B}
  \textbf{\bibinfo{volume}{87}}, \bibinfo{pages}{214504}
  (\bibinfo{year}{2013}).

\bibitem[{\citenamefont{Bud'ko et~al.}(1998)\citenamefont{Bud'ko, Canfield,
  Mielke, and Lacerda}}]{Bud1998Anisotropic}
\bibinfo{author}{\bibfnamefont{S.~L.} \bibnamefont{Bud'ko}},
  \bibinfo{author}{\bibfnamefont{P.~C.} \bibnamefont{Canfield}},
  \bibinfo{author}{\bibfnamefont{C.~H.} \bibnamefont{Mielke}},
  \bibnamefont{and} \bibinfo{author}{\bibfnamefont{A.~H.}
  \bibnamefont{Lacerda}}, \bibinfo{journal}{Phys. Rev. B}
  \textbf{\bibinfo{volume}{57}}, \bibinfo{pages}{13624} (\bibinfo{year}{1998}).

\bibitem[{\citenamefont{Goodrich et~al.}(2004)\citenamefont{Goodrich, Browne,
  Kurtz, Young, DiTusa, Adams, and Hall}}]{Goodrich2004deHaasvan}
\bibinfo{author}{\bibfnamefont{R.~G.} \bibnamefont{Goodrich}},
  \bibinfo{author}{\bibfnamefont{D.}~\bibnamefont{Browne}},
  \bibinfo{author}{\bibfnamefont{R.}~\bibnamefont{Kurtz}},
  \bibinfo{author}{\bibfnamefont{D.~P.} \bibnamefont{Young}},
  \bibinfo{author}{\bibfnamefont{J.~F.} \bibnamefont{DiTusa}},
  \bibinfo{author}{\bibfnamefont{P.~W.} \bibnamefont{Adams}}, \bibnamefont{and}
  \bibinfo{author}{\bibfnamefont{D.}~\bibnamefont{Hall}},
  \bibinfo{journal}{Phys. Rev. B} \textbf{\bibinfo{volume}{69}},
  \bibinfo{pages}{125114} (\bibinfo{year}{2004}).

\bibitem[{\citenamefont{Young et~al.}(2003)\citenamefont{Young, Goodrich,
  DiTusa, Guo, Adams, Chan, and Hall}}]{Young2003High}
\bibinfo{author}{\bibfnamefont{D.~P.} \bibnamefont{Young}},
  \bibinfo{author}{\bibfnamefont{R.~G.} \bibnamefont{Goodrich}},
  \bibinfo{author}{\bibfnamefont{J.~F.} \bibnamefont{DiTusa}},
  \bibinfo{author}{\bibfnamefont{S.}~\bibnamefont{Guo}},
  \bibinfo{author}{\bibfnamefont{P.~W.} \bibnamefont{Adams}},
  \bibinfo{author}{\bibfnamefont{J.~Y.} \bibnamefont{Chan}}, \bibnamefont{and}
  \bibinfo{author}{\bibfnamefont{D.}~\bibnamefont{Hall}},
  \bibinfo{journal}{Appl. Phys. Lett.} \textbf{\bibinfo{volume}{82}},
  \bibinfo{pages}{3713} (\bibinfo{year}{2003}).

\bibitem[{\citenamefont{Parish and Littlewood}(2003)}]{Parish2003Nonsaturating}
\bibinfo{author}{\bibfnamefont{M.~M.} \bibnamefont{Parish}} \bibnamefont{and}
  \bibinfo{author}{\bibfnamefont{P.~B.} \bibnamefont{Littlewood}},
  \bibinfo{journal}{Nature} \textbf{\bibinfo{volume}{426}},
  \bibinfo{pages}{162} (\bibinfo{year}{2003}).

\bibitem[{\citenamefont{Wang et~al.}(2012)\citenamefont{Wang, Du, Dou, and
  Zhang}}]{Wang2012Room}
\bibinfo{author}{\bibfnamefont{X.}~\bibnamefont{Wang}},
  \bibinfo{author}{\bibfnamefont{Y.}~\bibnamefont{Du}},
  \bibinfo{author}{\bibfnamefont{S.}~\bibnamefont{Dou}}, \bibnamefont{and}
  \bibinfo{author}{\bibfnamefont{C.}~\bibnamefont{Zhang}},
  \bibinfo{journal}{Phys. Rev. Lett.} \textbf{\bibinfo{volume}{108}},
  \bibinfo{pages}{266806} (\bibinfo{year}{2012}).

\bibitem[{\citenamefont{Dasoundhi et~al.}(2021)\citenamefont{Dasoundhi, Rajput,
  Kumar, and Lakhani}}]{Dasoundhi2021Extremely}
\bibinfo{author}{\bibfnamefont{M.~K.} \bibnamefont{Dasoundhi}},
  \bibinfo{author}{\bibfnamefont{I.}~\bibnamefont{Rajput}},
  \bibinfo{author}{\bibfnamefont{D.}~\bibnamefont{Kumar}}, \bibnamefont{and}
  \bibinfo{author}{\bibfnamefont{A.}~\bibnamefont{Lakhani}},
  \bibinfo{journal}{J. Phys D. Appl. Phys} \textbf{\bibinfo{volume}{54}},
  \bibinfo{pages}{195303} (\bibinfo{year}{2021}).

\bibitem[{\citenamefont{Qu et~al.}(2010)\citenamefont{Qu, Hor, Xiong, Cava, and
  Ong}}]{Qu2010Quantum}
\bibinfo{author}{\bibfnamefont{D.~X.} \bibnamefont{Qu}},
  \bibinfo{author}{\bibfnamefont{Y.~S.} \bibnamefont{Hor}},
  \bibinfo{author}{\bibfnamefont{J.}~\bibnamefont{Xiong}},
  \bibinfo{author}{\bibfnamefont{R.~J.} \bibnamefont{Cava}}, \bibnamefont{and}
  \bibinfo{author}{\bibfnamefont{N.~P.} \bibnamefont{Ong}},
  \bibinfo{journal}{Science} \textbf{\bibinfo{volume}{329}},
  \bibinfo{pages}{821} (\bibinfo{year}{2010}).

\bibitem[{\citenamefont{Ruszała et~al.}(2020)\citenamefont{Ruszała,
  Winiarski, and Samsel-Czekała}}]{Rusza2020DiracLike}
\bibinfo{author}{\bibfnamefont{P.}~\bibnamefont{Ruszała}},
  \bibinfo{author}{\bibfnamefont{M.~J.} \bibnamefont{Winiarski}},
  \bibnamefont{and}
  \bibinfo{author}{\bibfnamefont{M.}~\bibnamefont{Samsel-Czekała}},
  \bibinfo{journal}{Acta. Phys. Pol. A} \textbf{\bibinfo{volume}{138}},
  \bibinfo{pages}{748} (\bibinfo{year}{2020}).

\bibitem[{\citenamefont{Leonhardt et~al.}(2021)\citenamefont{Leonhardt,
  Hirschmann, Heinsdorf, Wu, Fabini, and Schnyder}}]{Leonhardt2021}
\bibinfo{author}{\bibfnamefont{A.}~\bibnamefont{Leonhardt}},
  \bibinfo{author}{\bibfnamefont{M.~M.} \bibnamefont{Hirschmann}},
  \bibinfo{author}{\bibfnamefont{N.}~\bibnamefont{Heinsdorf}},
  \bibinfo{author}{\bibfnamefont{X.}~\bibnamefont{Wu}},
  \bibinfo{author}{\bibfnamefont{D.~H.} \bibnamefont{Fabini}},
  \bibnamefont{and} \bibinfo{author}{\bibfnamefont{A.~P.}
  \bibnamefont{Schnyder}}, \bibinfo{journal}{Phys. Rev. Mater.}
  \textbf{\bibinfo{volume}{5}}, \bibinfo{pages}{124202} (\bibinfo{year}{2021}).

\bibitem[{\citenamefont{Abrikosov}(1998)}]{Abrikosov1998Quantum}
\bibinfo{author}{\bibfnamefont{A.~A.} \bibnamefont{Abrikosov}},
  \bibinfo{journal}{Phys. Rev. B} \textbf{\bibinfo{volume}{58}},
  \bibinfo{pages}{2788} (\bibinfo{year}{1998}).

\bibitem[{\citenamefont{Fischer et~al.}(2019)\citenamefont{Fischer, Roth, and
  Iversen}}]{Fischer2019Transport}
\bibinfo{author}{\bibfnamefont{K.~F.~F.} \bibnamefont{Fischer}},
  \bibinfo{author}{\bibfnamefont{N.}~\bibnamefont{Roth}}, \bibnamefont{and}
  \bibinfo{author}{\bibfnamefont{B.~B.} \bibnamefont{Iversen}},
  \bibinfo{journal}{J. Appl. Phys} \textbf{\bibinfo{volume}{125}},
  \bibinfo{pages}{045110} (\bibinfo{year}{2019}).

\bibitem[{SM()}]{SM}
\bibinfo{note}{See Supplemental Material for details of methods (calculations,
  ARPES and Single crystal synthesis and characterization, which includes
  Refs.~\cite{Leonhardt2021, Shao2019composite, Fischer2019Transport,
  Rusza2020DiracLike, Zhao2018Fermi, Pippard1989Magnetoresistance,
  Young2003High, Goodrich2004deHaasvan, Lifshits1958Theory}}.

\bibitem[{\citenamefont{Klemenz et~al.}(2019)\citenamefont{Klemenz, Lei, and
  Schoop}}]{Klemenz2019}
\bibinfo{author}{\bibfnamefont{S.}~\bibnamefont{Klemenz}},
  \bibinfo{author}{\bibfnamefont{S.}~\bibnamefont{Lei}}, \bibnamefont{and}
  \bibinfo{author}{\bibfnamefont{L.~M.} \bibnamefont{Schoop}},
  \bibinfo{journal}{Annu. Rev. Mater. Sci.} \textbf{\bibinfo{volume}{49}},
  \bibinfo{pages}{185} (\bibinfo{year}{2019}).

\bibitem[{\citenamefont{Liebowitz and Shevchik}(1978)}]{Liebowitz1978Free}
\bibinfo{author}{\bibfnamefont{D.}~\bibnamefont{Liebowitz}} \bibnamefont{and}
  \bibinfo{author}{\bibfnamefont{N.~J.} \bibnamefont{Shevchik}},
  \bibinfo{journal}{Phys. Rev. B} \textbf{\bibinfo{volume}{17}},
  \bibinfo{pages}{3825} (\bibinfo{year}{1978}).

\bibitem[{\citenamefont{Shao et~al.}(2019)\citenamefont{Shao, Wang, Chen, Lu,
  Gu, Sheng, Xing, and Sun}}]{Shao2019composite}
\bibinfo{author}{\bibfnamefont{D.}~\bibnamefont{Shao}},
  \bibinfo{author}{\bibfnamefont{H.}~\bibnamefont{Wang}},
  \bibinfo{author}{\bibfnamefont{T.}~\bibnamefont{Chen}},
  \bibinfo{author}{\bibfnamefont{P.}~\bibnamefont{Lu}},
  \bibinfo{author}{\bibfnamefont{Q.}~\bibnamefont{Gu}},
  \bibinfo{author}{\bibfnamefont{L.}~\bibnamefont{Sheng}},
  \bibinfo{author}{\bibfnamefont{D.}~\bibnamefont{Xing}}, \bibnamefont{and}
  \bibinfo{author}{\bibfnamefont{J.}~\bibnamefont{Sun}}, \bibinfo{journal}{npj
  Comput. Mater.} \textbf{\bibinfo{volume}{5}}, \bibinfo{pages}{53}
  (\bibinfo{year}{2019}).

\bibitem[{\citenamefont{Zhao et~al.}(2018)\citenamefont{Zhao, Xu, Zuo, Wu, Gao,
  and Zhu}}]{Zhao2018Fermi}
\bibinfo{author}{\bibfnamefont{L.}~\bibnamefont{Zhao}},
  \bibinfo{author}{\bibfnamefont{L.}~\bibnamefont{Xu}},
  \bibinfo{author}{\bibfnamefont{H.}~\bibnamefont{Zuo}},
  \bibinfo{author}{\bibfnamefont{X.}~\bibnamefont{Wu}},
  \bibinfo{author}{\bibfnamefont{G.}~\bibnamefont{Gao}}, \bibnamefont{and}
  \bibinfo{author}{\bibfnamefont{Z.}~\bibnamefont{Zhu}},
  \bibinfo{journal}{Phys. Rev. B} \textbf{\bibinfo{volume}{98}},
  \bibinfo{pages}{085137} (\bibinfo{year}{2018}).

\bibitem[{\citenamefont{Pippard}(1989)}]{Pippard1989Magnetoresistance}
\bibinfo{author}{\bibfnamefont{A.~B.} \bibnamefont{Pippard}},
  \emph{\bibinfo{title}{Magnetoresistance in Metal}}
  (\bibinfo{publisher}{Cambridge University Press},
  \bibinfo{address}{Cambridge, England}, \bibinfo{year}{1989}).

\bibitem[{\citenamefont{Lifshits and Kosevich}(1958)}]{Lifshits1958Theory}
\bibinfo{author}{\bibfnamefont{E.~M.} \bibnamefont{Lifshits}} \bibnamefont{and}
  \bibinfo{author}{\bibfnamefont{A.~M.} \bibnamefont{Kosevich}},
  \bibinfo{journal}{J. Phys. Chem. Solids} \textbf{\bibinfo{volume}{4}},
  \bibinfo{pages}{1} (\bibinfo{year}{1958}).

\end{thebibliography}

\end{document}

\title{Supplementary Material for\\ ``Multiple topological nodal structure in LaSb$_2$ with large linear magnetoresistance''}

\author{Y. X. Qiao}\altaffiliation{These authors contributed equally to this work}
\affiliation{State Key Laboratory of Functional Materials for Informatics, Shanghai Institute of Microsystem and Information Technology, Chinese Academy of Sciences, Shanghai 200050, China}
\affiliation{Center of Materials Science and Optoelectronics Engineering, University of Chinese Academy of Sciences, Beijing 100049, China}

\author{Z. C. Tao}\altaffiliation{These authors contributed equally to this work}
\affiliation{School of Physical Science and Technology, ShanghaiTech University, Shanghai 201210, China}

\author{F. Y. Wang}\altaffiliation{These authors contributed equally to this work}
\affiliation{National Laboratory of Solid State Microstructures, School of Physics, Nanjing University, Nanjing 210093, China}
\affiliation{Collaborative Innovation Center of Advanced Microstructures, Nanjing University, Nanjing 210093, China}

\author{Huaiqiang Wang}
\email{hqwang@nju.edu.cn}
\affiliation{National Laboratory of Solid State Microstructures, School of Physics, Nanjing University, Nanjing 210093, China}
\affiliation{Collaborative Innovation Center of Advanced Microstructures, Nanjing University, Nanjing 210093, China}

\author{Z. C. Jiang}
\affiliation{State Key Laboratory of Functional Materials for Informatics, Shanghai Institute of Microsystem and Information Technology, Chinese Academy of Sciences, Shanghai 200050, China}

\author{Z. T. Liu}
\affiliation{State Key Laboratory of Functional Materials for Informatics, Shanghai Institute of Microsystem and Information Technology, Chinese Academy of Sciences, Shanghai 200050, China}

\author{Soohyun Cho}
\affiliation{State Key Laboratory of Functional Materials for Informatics, Shanghai Institute of Microsystem and Information Technology, Chinese Academy of Sciences, Shanghai 200050, China}

\author{F. Y. Zhang}
\affiliation{State Key Laboratory of Functional Materials for Informatics, Shanghai Institute of Microsystem and Information Technology, Chinese Academy of Sciences, Shanghai 200050, China}
\affiliation{Shenzhen Institute for Quantum Science and Technology and Department of Physics, Southern University of Science and Technology, Shenzhen 518055, China}

\author{Q. K. Meng}
\affiliation{Wuhan National High Magnetic Field Center and School of Physics, Huazhong University of Science and Technology, Wuhan 430074, China}

\author{W. Xia}
\affiliation{School of Physical Science and Technology, ShanghaiTech University, Shanghai 201210, China}

\author{Y. C. Yang}
\affiliation{State Key Laboratory of Functional Materials for Informatics, Shanghai Institute of Microsystem and Information Technology, Chinese Academy of Sciences, Shanghai 200050, China}

\author{Z. Huang}
\affiliation{State Key Laboratory of Functional Materials for Informatics, Shanghai Institute of Microsystem and Information Technology, Chinese Academy of Sciences, Shanghai 200050, China}
\affiliation{School of Physical Science and Technology, ShanghaiTech University, Shanghai 201210, China}

\author{J. S. Liu}
\affiliation{State Key Laboratory of Functional Materials for Informatics, Shanghai Institute of Microsystem and Information Technology, Chinese Academy of Sciences, Shanghai 200050, China}

\author{Z. H. Liu}
\affiliation{State Key Laboratory of Functional Materials for Informatics, Shanghai Institute of Microsystem and Information Technology, Chinese Academy of Sciences, Shanghai 200050, China}

\author{Z. W. Zhu}
\affiliation{Wuhan National High Magnetic Field Center and School of Physics, Huazhong University of Science and Technology, Wuhan 430074, China}

\author{S. Qiao}
\affiliation{State Key Laboratory of Functional Materials for Informatics, Shanghai Institute of Microsystem and Information Technology, Chinese Academy of Sciences, Shanghai 200050, China}
\affiliation{Center of Materials Science and Optoelectronics Engineering, University of Chinese Academy of Sciences, Beijing 100049, China}
\affiliation{School of Physical Science and Technology, ShanghaiTech University, Shanghai 201210, China}

\author{Y. F. Guo}
\email{guoyf@shanghaitech.edu.cn}
\affiliation{School of Physical Science and Technology, ShanghaiTech University, Shanghai 201210, China}
\affiliation{ShanghaiTech Laboratory for Topological Physics, ShanghaiTech University, Shanghai 201210, China}

\author{H. J. Zhang}
\email{zhanghj@nju.edu.cn}
\affiliation{National Laboratory of Solid State Microstructures, School of Physics, Nanjing University, Nanjing 210093, China}
\affiliation{Collaborative Innovation Center of Advanced Microstructures, Nanjing University, Nanjing 210093, China}

\author{D. W. Shen}
\email{dwshen@mail.sim.ac.cn}
\affiliation{State Key Laboratory of Functional Materials for Informatics, Shanghai Institute of Microsystem and Information Technology, Chinese Academy of Sciences, Shanghai 200050, China}
\affiliation{Center of Materials Science and Optoelectronics Engineering, University of Chinese Academy of Sciences, Beijing 100049, China}

\maketitle

\tableofcontents
\newpage
\section{Methods}
\subsection{Sample growth and characterizations}
High-quality LaSb$_2$ single crystals were grown by using the self-flux method. Starting materials La (99.5\%, Macklin) and Sb (99.99\%, Aladdin) blocks were mixed in a molar ratio of 1:15 and placed into an alumina crucible, which was then sealed into a quartz tube in vacuum. The assembly was heated in a furnace up to 1050$^{\circ}$C within 10 hrs, kept at that temperature for 10 hrs, and then slowly cooled down to 700$^{\circ}$C at a temperature decreasing rate of 2$^{\circ}$C/h. The excess Sb was removed at this temperature by quickly placing the assembly into a high-speed centrifuge, and black crystals with shining surface in a typical dimension of $5\times2.5\times0.2$ $mm^{3}$ were obtained. The crystallographic phase and quality examinations of LaSb$_2$ were performed on a Bruker D8 single-crystal X-ray diffractometer (SXRD) with Mo $K_{\alpha1}(\lambda=0.71073\AA)$ at 298 K. The powder XRD was carried on a Bruker D8 Advance powder X-ray diffractometer (PXRD) with Cu $K_{\alpha1}(\lambda=1.54184\AA)$ at 298 K. 

\subsection{ARPES experiments}
ARPES measurements were performed at 03U and “Dreamline” beamline of the Shanghai Synchrotron Radiation Facility (SSRF) with a Scienta Omicron DA30L analyzer. All samples were cleaved under a vacuum better than $8 \times 10^{-11} $ Torr. During measurements, the temperature was kept at 24 K and 210 K. 

\subsection{ \emph{ab initio} calculations}
DFT calculations were performed in the Vienna Ab initio Simulation Package(VASP) by implementing projector-augmented wave pseudopotential with exchange-correlation functional of generalized gradient approximation of Perdew, Burke, and Ernzerhof. Besides, calculations without/with SOC were performed to further discuss the special band structure of LaSb$_2$ (see Fig. S1), convergence criteria for energy and force are set to $10^{-7}$ eV and $10^{-5}$ eV respectively. We adopted $10 \times 10 \times 4$ k mesh for the first Brillouin zone. Maximally localized Waninger functions based on La $5d$ and Sb $5p$ orbitals were generated and then using WannierTools to obtain the Fermi surface and surface state spectrum.

\newpage
\section{Symmetry analysis of topological nodal states in LaSb$_2$}

\begin{table}[htbp]
	\begin{center}
		\caption{High-symmetry points and the corresponding symmetry operations, where $k_x$, $k_y$ and $k_z$ are given in units of $2/a$, $2/b$, and $1/c$, and $\xi\equiv (1+a^2/b^2)/2$. }
		\label{table I}
		{\begin{tabular}{c|c}
			\hline\hline
			High symmetry points & Symmetry operations\\\hline
			$\Gamma(0,0,0)$, $Z(0,0,\pi)$, $X(\pi,0,0)$, $A(\pi,0,\pi)$ & $\left\{E, I, M_{x}, G_{y}, G_{z}, C_{2 x}, S_{2 y}, S_{2 z}\right\}$\\\hline
			$R(\pi/2,\pi/2,\pi/2)$, $S(\pi/2,\pi/2,0)$ & $\left\{E, I, G_{z}, S_{2 z}\right\}$\\\hline
			$Y(0,\xi\pi,0)$, $T(0,\xi\pi,\pi)$, $Y_1(\pi,(1-\xi)\pi,0)$, $A_1(\pi,(1-\xi)\pi,\pi)$ & $\left\{E,  M_{x}, G_{z}, S_{2 y}\right\}$\\\hline\hline
		\end{tabular}}
	\end{center}
\end{table}

\begin{figure}
 \centering
  \includegraphics[width=6in]{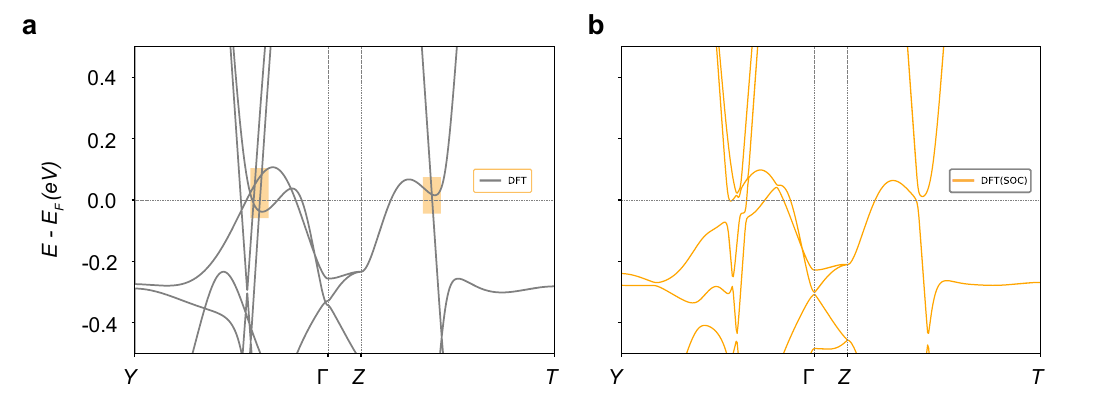}\\
  \caption{DFT calculations of the band structure without and with SOC along high-symmetry lines.}\label{fig.S1}
\end{figure}

In this section, we will provide the detailed symmetry analysis of various nodal states, especially symmetry-enforced crossings in LaSb$_2$ with and without spin-orbit coupling (SOC)~\cite{Leonhardt2021}. LaSb$_2$ crystalizes in the nonsymmorphic space group $Cmce$ (No. 64), with the relevant symmetry operations given as
\begin{equation}
\tag{S1}
\begin{aligned}
E=\{1\ |\  0,0,0\}, \ M_x=\{m_{100}\ |\  0,0,0\}, \ G_y=\{m_{010}\ |\  0,1/2,1/2\}, \ G_z=\{m_{001}\ |\  0,1/2, 1/2\} ,\\ 
I=\{-1\ |\  0,0,0\}, \ C_{2x}=\{2_{100}\ |\  0,0,0\}, \ S_{2y}=\{2_{010}\ |\  0,1/2,1/2\},\ S_{2z}=\{2_{001}\ |\  0,1/2,1/2\},
\end{aligned}
\end{equation}
where $G_y$ and $G_z$ are glide mirror symmetries, and $S_{2y}$ and $S_{2z}$ are screw rotation symmetries. As a base-centered orthorhombic lattice, the primitive lattice vectors of LaSb$_2$ are given by 
\begin{equation}
\tag{S2}
\mathbf{a}_{1}=\left(\frac{a}{2}, -\frac{b}{2}, 0\right), \ \ \mathbf{a}_{2}=\left(\frac{a}{2}, \frac{b}{2}, 0\right), \ \ \mathbf{a}_{3}=\left(0,0, c\right),
\end{equation} 
where $a,b$ and $c$ are the lattice parameters in the conventional basis. The reciprocal lattice vectors are then given by
\begin{equation}
\tag{S3}
\mathbf{b_1}=(\frac{2\pi}{a},-\frac{2\pi}{b},0), \ \ \mathbf{b_2}=(\frac{2\pi}{a},\frac{2\pi}{b},0),\ \ \mathbf{b_3}=(0,0,\frac{2\pi}{c}).
\end{equation}
Taking the relation $a>b$ into consideration, the Brillouin zone is then schematically in Fig. S2 (a), where corresponding high symmetry points have been explicitly labelled. For later references, here, we list relevant high-symmetry points,  high symmetry lines and high-symmetry planes in Tables. I, II and III, respectively.

\begin{table}[htbp]
	\begin{center}
		\caption{High-symmetry lines and the corresponding symmetry operations, where $k_x$, $k_y$ and $k_z$ are given in units of $2/a$, $2/b$, and $1/c$, and $\xi\equiv (1+a^2/b^2)/2$. }
		\label{table II}
		{\begin{tabular}{c|c}
			\hline\hline
			High symmetry lines& Symmetry operations\\\hline
			$Z-T(0, y, \pi) \quad \Gamma-Y(0, y, 0) \quad A-A_{1}(\pi, y, \pi) \quad X-Y_{1}(\pi, y, 0)$ & $\left\{E, M_{x},  G_{z}, S_{2 y}\right\}$\\\hline
			$\Gamma-Z(0,0, z) \quad X-A(\pi, 0, z)$ & $\left\{E, M_x, G_{y}, S_{2 z}\right\}$\\\hline
			$\Gamma-X(x, 0,0)$ $Z-A(x, 0, \pi)$ & $\left\{E,  G_y, G_z, C_{2x}\right\}$\\\hline
			$Y-T(0, \xi\pi, z) \quad Y_{1}-A_{1}(\pi, (1-\xi)\pi, z)$ & $\left\{E,  M_{x}\right\}$\\\hline
			$R-S(\pi/2, \pi/2, z)$ & $\left\{E,  S_{2z}\right\}$\\\hline\hline
		\end{tabular}}
	\end{center}
\end{table}

\begin{figure}
 \centering
  \includegraphics[width=6in]{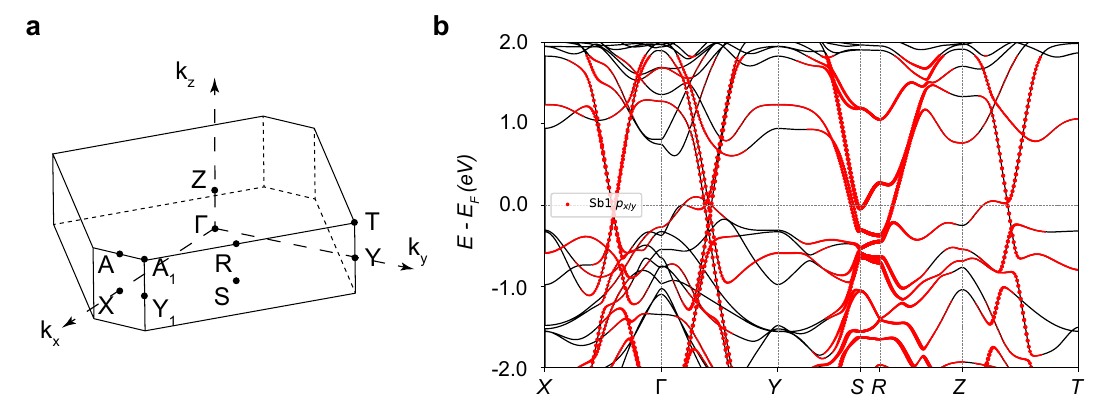}\\
  \caption{a. Schematic illustration of the Brillouin zone of LaSb$_2$ in the $Cmce$ space group with $a>b$. b. Calculated fatband structures withou SOC, which are dominated by $p_{x/y}$ orbitals from $Sb1$ atoms.}
  \label{fig.S2}
\end{figure}

\begin{table}[htbp]
	\begin{center}
		\caption{High-symmetry planes and the corresponding symmetry operations, where $k_x$, $k_y$ and $k_z$ are given in units of $2/a$, $2/b$, and $1/c$. }
		\label{table III}
		{\begin{tabular}{c|c}
			\hline\hline
			High symmetry planes& Symmetry operations\\\hline
			$\ k_x=\pi$,  $k_x=0$ & $\left\{E, M_x\right\}$\\\hline
			$\ k_y=\pi$,  $k_y=0$ & $\left\{E, G_{y}\right\}$\\\hline
			$\ k_z=\pi$,  $k_z=0$ & $\left\{E, G_{z}\right\}$\\\hline\hline
		\end{tabular}}
	\end{center}
\end{table}

\subsection{Topological nodal phenomena without SOC}
\subsubsection{Symmetry-enforced nodal surface in the $k_{z}=\pi$ plane}
In the $k_z=\pi$ plane, there exists a combined antiunitary symmetry $S_{2z}\Theta$ constructed by the screw rotation $S_{2z}$ and time-reversal operator $\Theta$. This operator satisfies
\begin{equation}
\tag{S4}
(S_{2z}\Theta)^2=t_{001}=e^{-ick_z }=-1,
\end{equation}
where $t_{001}$ represents a translation of $c$ along the $z$-direction. As a result, Kramers degeneracy is enforced for all momenta in the $k_z=\pi$ plane, leading to the symmetry-enforced doubly degenerate (excluding spin degeneracy) nodal surface states in this plane.

\subsubsection{Symmetry-enforced nodal line along $R-S$}
Along the $R-S$ line, where the combined antiunitary symmetry $G_z\Theta$ is preserved, we have 
\begin{equation}
\tag{S5}
(G_z\Theta)^2=t_{010}=e^{-ibk_y^{RS}}=-1.
\end{equation}
Similar to above analysis, this relation ensures Kramers degeneracy along $R-S$ line and results in doubly degenerate bands.

\subsection{Topological nodal phenomena with SOC}
\subsubsection{Symmetry-enforced fourfold degenerate nodal line along $Z-T$ and $A-A_1$}
Along the $Z-T$ ($A-A_1$) line, $G_y\Theta$, $S_{2y}$, $G_z$, and $P\Theta$ ($P$ means inversion) symmetries are preserved.  We will show that for any state $|u\rangle$, the four states $|u\rangle,P\Theta|u\rangle,G_y\Theta|u\rangle,S_{2y}|u\rangle$ are orthogonal to each other and form a degenerate quartet~\cite{Shao2019composite}. We assume that $G_z|u\rangle=g_z|u\rangle$. In the spinful case, we have
\begin{equation}
\tag{S6}
G_z^2=\bar{E}t_{010}=-e^{-ibk_y}\rightarrow g_z=\pm ie^{-ibk_y/2}.
\end{equation}
Since
\begin{equation}
\tag{S7}
G_zG_y=-t_{01\bar{1}}G_yG_z=-e^{-ibk_y+ick_z}G_yG_z=e^{-ibk_y}G_yG_z,
\end{equation}
we obtain
\begin{equation}
\tag{S8}
G_zG_y\Theta|u\rangle=e^{-ibk_y}G_y\Theta G_{z}|u\rangle=-g_zG_y\Theta|u\rangle,
\end{equation}
which means $G_y\Theta|u\rangle$ has opposite $G_z$ eigenvalues to that of $|u\rangle$. In addition, for the state $P\Theta|u\rangle$, because of the condition $(P\Theta)^2=-1$, it is a Kramers partner of $|u\rangle$ and they are mutually orthogonal. However, due to
\begin{equation}
\tag{S9}
G_zP=t_{011}PG_z=e^{-ibk_y-ick_z^{ZT}}PG_z=-e^{-ibk_y}PG_z,
\end{equation}
we have
\begin{equation}
\tag{S10}
G_zP\Theta|u\rangle=-e^{-ibk_y}P\Theta G_z|u\rangle=g_zP\Theta|u\rangle,
\end{equation}
indicating that $P\Theta |u\rangle$ and $|u\rangle$ have the same $G_z$ eigenvaules, and are thus orthogonal to $G_y\Theta|u\rangle$. Finally, the $P\Theta$ partner of $G_y\Theta|u\rangle$ will give rise to another orthogonal state $S_{2y}|u\rangle$. Consequently, all the bands along $Z-T$ ($A-A_1$) are fourfold degenerate nodal lines. 

\subsubsection{Symmetry-enforced fourfold degenerate Dirac point at S}
At the $S$ $(\pi/a,\pi/b,0)$ point, $S_{2z}$, $P$, $P\Theta$, $\Theta$ symmetries are preserved, and it can be shown that the four states $|u\rangle$, $P|u\rangle$, $P\Theta|u\rangle$, $\Theta|u\rangle$ are degenerate and mutually orthogonal. Assume that $S_{2z}|u\rangle=\lambda_z|u\rangle$. First, according to the relation 
\begin{equation}
\tag{S11}
S_{2z}^2=\bar{E}t_{001}=-e^{-ick_z},
\end{equation}
we get $\lambda_{z}=\pm i e^{-ick_z/2}=\pm i$ at S point, which is purely imaginary. As result, $\Theta|u\rangle$ has opposite eigenvalue of $\lambda_z$ and is orthogonal to $|u\rangle$. Moreover, based on the relation
\begin{equation}
\tag{S12}
S_{2z}P=t_{011}PS_{2z}=e^{-ibk_y-ick_z}PS_{2z},
\end{equation}
at S point, for the state $P|u\rangle$, we obtain 
\begin{equation}
\tag{S13}
S_{2z}P|u\rangle=-P S_{2z}|u\rangle=-\lambda_z P\Theta|u\rangle,
\end{equation}
and for the orthogonal $P\Theta$ partner of $|u\rangle$,
\begin{equation}
\tag{S14}
S_{2z}P\Theta|u\rangle=-P\Theta S_{2z}|u\rangle=\lambda_z P\Theta|u\rangle.
\end{equation}
The $S_{2z}$ eigenvalues of the four states \{$|u\rangle$, $P|u\rangle$, $P\Theta|u\rangle$, $\Theta|u\rangle$\} are then labelled as \{$\lambda_z,-\lambda_z,\lambda_z,-\lambda_z$\}, which indicates that they form a degenerate quartet, and generate a fourfold degenerate point at S.

\newpage
\section{Angle-resolved photoemission spectroscopy}
%overlapped to demonstrate the movement of the Dirac point. 
Fig. S3 (b) plots the measured bands along representative cuts C1, which are perpendicular to the $k_z = \pi$, as indicated in Fig. S3 (a). The bands along cuts C1 exhibit several Dirac-like crossings on every $Z$ point and forming nodal surface at $k_z = \pi$, which shows nice overall consistency between experiments and \emph{ab initio} calculations.
Besides, we present the Fermi surface (FS) map in Fig. S3 (d), which clearly suggests the nearly square BZ and existing many electron and hole pockets cross the fermi level, further proving it has similar characters compared with ZrSiS. As the result of weak dispersion of $k_z$ dependence, constant-energy surfaces at different photon energies are similarly.

\begin{figure}
 \centering
  %Requires \usepackage{graphicx}
  \includegraphics{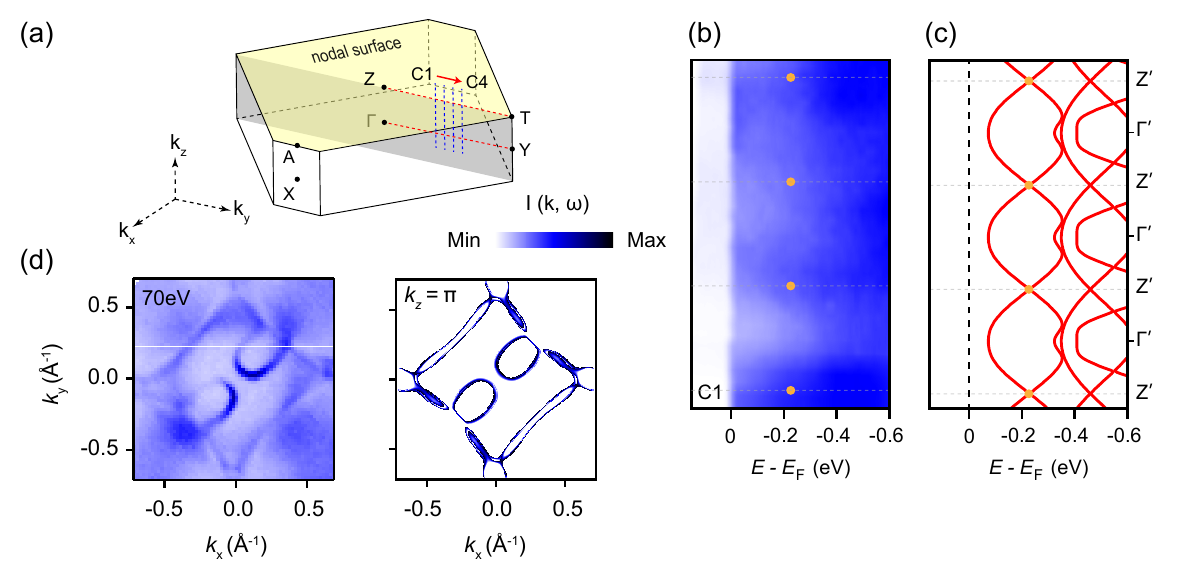}\\
  \caption{(a) Schematic of momentum locations of cuts C1 to C4, $Z-T$ and $\Gamma-Y$ in the bulk BZ. (b) Band dispersions along cuts C1, the orange dots represent Dirac points at $k_z = \pi$. (c) The corresponding calculated band structure of (b). (d) Comparision between the experimental and bulk calculated constant-energy surface.}\label{fig.S3}
\end{figure}

\newpage
\section{Crystal structure and transportation results}

\subsection{Single crystal characterization}

The phase and quality examinations of LaSb$_2$ were performed by XRD, and diffraction pattern could be satisfyingly indexed on the basis of the pyrite-type structure in the nonsymmorphic orthorhombic $Cmca$ (No. 64) space group [Fig.~S4 (a)]. The extracted lattice parameters are $a = 6.314$ $\AA$, $b = 6.175$ $\AA$ and $c = 18.671$ $\AA$, respectively, consistent with those in previous literature~\cite{Fischer2019Transport}. We note that our single crystals show flat and shining surfaces after cleavage in the air, as illustrated in the inset of Fig.~S4 (a).

\subsection{Magnetotransport measurements and Quantum oscillations}

The low-field magnetotransport measurements, including the resistivity and Hall effect measurements, were carried out using a standard Hall bar geometry in a commercial DynaCool physical properties measurement system (PPMS) from Quantum Design. High-field measurements were performed at Hefei High Magnetic Field Laboratory, Chinese Academy of Sciences, and Wuhan National High Magnetic Field Center.

\subsubsection{Unsaturated linear magnetoresistance}  
The LMR of LaSb$_2$ at 1.8~K is reproduced in Fig.~S4 (b), which exhibits linear behavior and does not saturate with increase of magnetic field up to 32~T. The Fermi surface from DFT calculations is drawn in Fig. S4 (c). Previous theoretical studies\cite{Rusza2020DiracLike} suggested the absence of carrier compensation in LaSb$_2$. To experimentally confirm this, we conducted magnetic field dependent Hall conductivity measurements and the result is presented as the inset of Fig. S4 (d). To quantitatively estimate the densities and mobilities of the carriers, the Hall conductivity, expressed as  $\sigma_{xy}=-\rho_{xy}/(\rho_{xy}^{2}+\rho_{xx}^{2})$, is fitted by employing the semiclassical two-band model\cite{Pippard1989Magnetoresistance,Zhao2018Fermi}
\begin{equation}
\tag{S15}
	\sigma_{xy}=[\dfrac{n_{h}\mu_{h}^{2}}{1+(\mu_{h}B)^{2}}-\dfrac{n_{e}\mu_{e}^{2}}{1+(\mu_{e}B)^{2}}]eB
\end{equation} 
where $ n_{e} $ ($n_{h} $) denotes the carrier density for the electron (hole) and $ \mu_{e} $ ($ \mu_{h} $) represents the mobility of the electron (hole). The nice fitting results at 2 K are presented in the inset of Fig. S4 (g), yielding temperature dependences of densities and mobilities for electron and hole carriers as shown in Figs. S4 (d) and (g), respectively. The carriers densities are $n_{h}=1.55\times 10^{20}{\rm cm}^{-3}$ and $n_{e}=3.48\times 10^{19}{\rm cm}^{-3}$, revealing that the hole and electron carriers are in fact uncompensated. Though the densities of hole carriers is about one order of magnitude larger than that of electron carriers, the mobility of hole carriers is about one order of magnitude smaller than that of electron carriers. Both $ n_{h} $ and $ n_{e} $ decrease monotonically with decreasing temperatures. The hole carriers are the dominant carrier between $ T=2-200 $ K, consistent with the result reported earlier\cite{Young2003High}. The results exclude the carriers compensation mechanism for the LMR. Instead, we suggest that the quantum limit of Dirac-like nodal fermions with high Fermi velocities in high fields might play a crucial role.  

\begin{figure*}
 \centering
  %Requires \usepackage{graphicx}
  \includegraphics{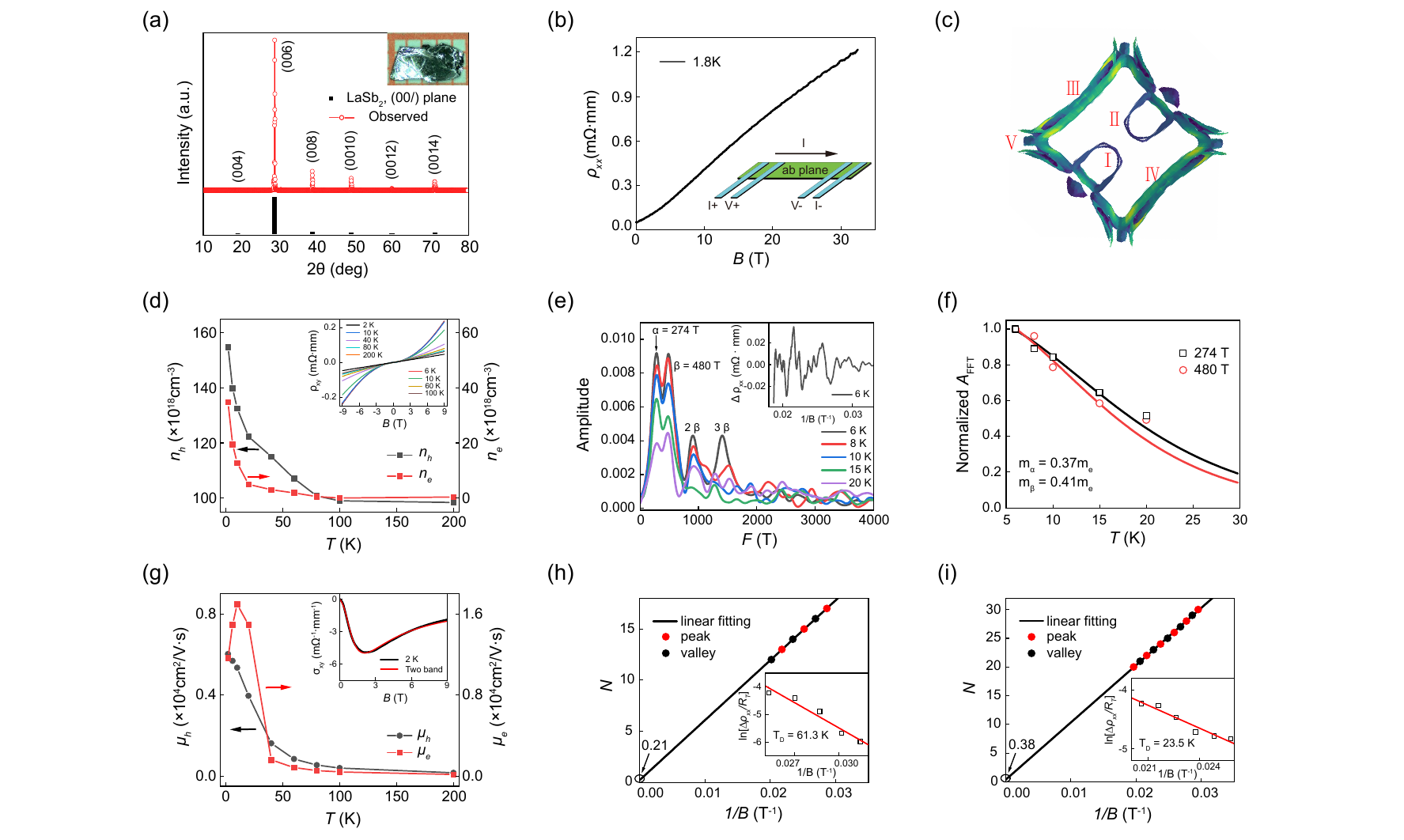}\\
  \caption{(a) Powder x-ray-diffraction (XRD) pattern of LaSb$_2$ (observed pattern, red line; calculation with pyrite structure type [$Cmca$, space group 64], black line). The inset shows the photo of a typical LaSb$_2$ single crystal. (b) Magnetic field dependent MR of LaSb$_{2}$ single crystal for $H//c$-axis at 1.8 K. The sketch for the four-probe measurement configuration is shown by the inset. (c) The calculated bulk FSs of LaSb$_2$. Roman numerals I and II are hole-like pockets while III–V are electron-like pockets. (d) Temperature dependence of carrier densities. Inset shows Hall conductivity versus magnetic field $B$ at different temperatures. (e) FFT spectra of $\Delta\rho_{xx}$. Inset shows the SdH oscillatory component as a function of 1/$B$ after subtracting background at 6 K. (f) Temperature dependence of relative amplitudes of the FFT spectra for the fundamental frequencies of 480 T (red circles) and 274 T (black squares). The solid lines denote the fitting result by using the LK formula. (g) Temperature dependence of carrier mobilities. Inset shows Hall conductivity at 2 K and the red line denotes the fitting results by the two-band model. (h) and (i) The Landau level fan diagrams for the fundamental frequencies of $\alpha$ and $\beta$, respectively. The corresponding insets show the Dingle fitting results.}\label{fig.S4}
\end{figure*}

\subsubsection{Quantum oscillations} 
The MR of LaSb$_{2} $ display clear quantum oscillations at $B$ \textgreater  30 T below 10 K, which could be seen in the Fig. S4 (e), consistent with those reported earlier\cite{Goodrich2004deHaasvan}. After carefully subtracting a sixth-order polynomial background, the $ \Delta\rho_{xx} $ exhibits striking Shubnikov–de Haas (SdH) oscillations. The SdH oscillations at 6 K against the reciprocal magnetic field 1/$ B $ are plotted by the inset of Fig. S4 (e), which could be well described by the Lifshitz-Kosevich (LK) formula\cite{Lifshits1958Theory} \par
\begin{equation}\tag{S16}
	\Delta\rho\propto R_{T}R_{D}\cos[2\pi(\dfrac{F}{B}+\gamma-\delta+\varphi)]
\end{equation}
where $ R_{T}= 2\pi^{2}k_{B}T/\hbar\omega_{c}/\sinh(2\pi^{2}k_{B}T/\hbar\omega_{c}) $, $ R_{D}=\exp(-2\pi^{2}k_{B}T_{D}/\hbar\omega_{c}) $, $ k_{B} $ is the Boltzmann constant, $ \hbar $ is Planck' constant, $ F $ is the oscillation's frequency, $\varphi$ is the phase shift, $\omega_{c}=eB/m^{\ast}$ is the cyclotron frequency, with $ m^{\ast} $ denoting the effective cyclotron mass, and $ T_{D} $ is the Dingle temperature. The fast Fourier transform(FFT) spectra of the SdH oscillations, shown in Fig. S4 (e), disclose multiple fundamental frequencies which are labeled as $ \alpha $ (274 T), $ \beta $ (480 T), $ \beta^{*} $ (960 T), $ \beta^{**} $ (1440 T), where $ F_{\beta^{*}}=2F_{\beta} $ and  $ F_{\beta^{**}}=3F_{\beta} $ are the harmonic value of $ F_{\beta} $, indicating two pockets across the Fermi level $ E_{F} $. The corresponding external cross-sectional areas of the Fermi surface can be determined by the Onsager relation $ F=(\hbar/2\pi e)A $. The effective cyclotron mass $ m^{\ast} $ at $ E_{F} $ could be obtained by fitting the temperature dependence of the FFT amplitude by the temperature damping factor of the LK equation expressed as $ R_{T} $, as is shown in Fig. S4 (f), giving $ m^{\ast}_{\alpha}=0.37m_{e} $ and $ m^{\ast}_{\beta}=0.41m_{e} $, where $ m_{e} $ denotes the free-electron mass. Fitting to the field dependent amplitudes of the quantum oscillations at 6 K, as shown in the insets of Figs. S4 (h) and (i), gives the Dingle temperatures $ T_{D} $ = 61.3 K and 23.5 K for bands $ \alpha $ and $ \beta $, respectively. The derived parameters are summarized in Table V.

To explore the topological nature of the electronic bands, the Landau level (LL) index fan diagram is established. The pseudospin rotation under a magnetic field in a Dirac/Weyl system will produce a nontrivial $ \varphi_{B} $ that could be accessed from the LL index fan diagram or a direct fit to the SdH oscillations by using the LK formula. The phase shift is generally a sum expressed as $ 1/2-\varphi_{B}/2\pi $ and $\delta$ represents the dimension-dependent correction to the phase shift, which is 0 for two-dimensional system and $ \pm 1/8$ for 3D system for nontrivial Berry phase, where the sign depends on the type of charge carriers and the kind of cross-section extremum. To provide more evidence for the nontrivial topological state in LaSb$ _{2} $, the Berry phase $ \varphi_{B} $ was examined. 
The LL index phase diagrams of $ \alpha $ and $ \beta $ are shown in Figs. S4 (h) and (i), in which the $ \Delta\rho_{xx} $ valley positions (black circles) in 1/$ B $ were assigned to be integer indices and the $ \Delta\rho_{xx} $ peak positions (red circles) were assigned to be half-integer indices. All the points almost fall on a straight line, thus allowing a linear fitting that gives the intercepts of 0.21 and 0.38 corresponding to $ \alpha $ and $ \beta $, respectively. The derived Berry phases $ \varphi_{B} $ are 0.17$ \pi $ and 1.01$ \pi $ for the electronic pocket $ \alpha $ and hole pocket $ \beta $. The Berry phase for the $ \beta $ frequency is close to $ \pi $, thus strongly indicating a nontrivial topological state in LaSb$ _{2} $.\par

\begin{table*}[htb]
	\begin{center}
		\caption{Parameters derived from SdH oscillations for LaSb$ _{2} $. Here $ k_{\rm F} $ is the Fermi wave vector, $ v_{\rm F} $ denotes the Fermi velocity, $ \mu $ represents the carrier mobility, $ \tau $ is the relaxation time, and $\varphi_{B}  $ is the Berry phase.}
		\label{table IV}
		\setlength{\tabcolsep}{3mm}{\begin{tabular}{l|l|l|l|l|l|l|l|l|l}
			\hline\hline
			$ F $(T) & $ A_{\rm F}(\rm nm^{-2}) $ & $ k_{\rm F} $(nm$ ^{-1} $) & $ v_{\rm F} $(m/s) & $ E_{\rm F} $(meV) & $ m^{*}/m_{0} $ & $ T_{\rm D} $(K)&$ \tau $(s)&$\mu$(cm$ ^{2} $/Vs)&$\varphi_{B}  $\\\hline
			274&2.62&0.91&3.84$ \times 10^{6}$&0.17&0.37&61.3&2$ \times 10^{-14} $&93.7&0.173
			\\480&4.58&1.21&3.35$ \times 10^{6}$&0.26&0.41&23.5&5$\times 10^{-14}$&217.7&1.01\\\hline\hline
		\end{tabular}}
	\end{center}
\end{table*}

%%%%%%%%%%%%%%%%%%%%%%%%%%%%%%%%%%%%%%%%%%%%%%%%%%%%%%%%%%%%%%%%%%%%%%%%%%%%%%%%%%%%%%%%%%
\bibliographystyle{apsrev4-2}

\bibliography{suppleref}

%\begin{thebibliography}{99}

%\end{thebibliography}